\documentclass[%
 reprint,
superscriptaddress,
 amsmath,amssymb,
 aps,
pre,
]{revtex4-2}

\usepackage{siunitx}
\usepackage{graphicx}
\usepackage{dcolumn}
\usepackage{bm}
\usepackage[export]{adjustbox}
\usepackage{caption}
\usepackage{subcaption}

\DeclareMathOperator\erfc{\textrm{erfc}}
\usepackage{soul}
\usepackage{xcolor}
\sethlcolor{red}

\newcommand{\HL}[1]{\textcolor{black}{#1}}

\begin{document}

\preprint{APS/123-QED}

\title{Effect of quenched heterogeneity on creep lifetimes of disordered materials}

\author{Juan Carlos Verano-Espitia}
\email{juan-carlos.verano-espitia@univ-grenoble-alpes.fr}
\affiliation{Univ. Grenoble Alpes, CNRS, ISTerre, 38000 Grenoble, France}
\affiliation{Department of Applied Physics, Aalto University, P.O. Box 15600, 00076 Aalto, Espoo, Finland}
\author{Jérôme Weiss}
\affiliation{Univ. Grenoble Alpes, CNRS, ISTerre, 38000 Grenoble, France}
\author{David Amitrano}
\affiliation{Univ. Grenoble Alpes, CNRS, ISTerre, 38000 Grenoble, France}
\author{Tero Mäkinen}
\affiliation{Department of Applied Physics, Aalto University, P.O. Box 15600, 00076 Aalto, Espoo, Finland}
\author{Mikko Alava}
\affiliation{Department of Applied Physics, Aalto University, P.O. Box 15600, 00076 Aalto, Espoo, Finland}
\affiliation{NOMATEN Centre of Excellence, National Centre for Nuclear Research, 05-400 Otwock-\'{S}wierk, Poland}

\date{\today}

\begin{abstract}

We revisit the problem of describing creep in heterogeneous materials by an effective temperature by considering more realistic (and complex) non-mean-field elastic redistribution kernels. We show first, from theoretical considerations, that, if elastic stress redistribution and memory effects are neglected, the average creep failure time follows an Arrhenius expression with an effective temperature explicitly increasing with the quenched heterogeneity. Using a thermally activated progressive damage model of compressive failure, we show that this holds true when taking into account elastic interactions and memory effects, however with an effective temperature $T_{eff}$ depending as well on the nature of the (non-democratic) elastic interaction kernel. We observe that the variability of creep lifetimes, for given external conditions of load and temperature, is roughly proportional to the mean lifetime, therefore depends as well on $T$, on quenched heterogeneity, and the elastic kernel. Finally, we discuss the implications of this effective temperature effect on the interpretation of macroscopic creep tests to estimate an activation volume at the microscale.  

\end{abstract}

\maketitle

\section{INTRODUCTION}

Creep, i.e.~the time-dependent deformation of materials under a constant load, as well as associated phenomena such as stress relaxation under a constant strain, are of great importance in many fields, such as material \citep{kassner2015fundamentals,franccois2012mechanical} and civil \citep{bazant1982creep} engineering, soft matter physics \citep{cipelletti2020microscopic, leocmach2014creep, rosti2010fluctuations, koivisto2016predicting, makinen2020scale, pournajar2023failure, miranda2024fractional} or geophysics and rocks mechanics \citep{duval2010creep,brantut2013,savage2005postseismic}. While submitted to an external load constant in time, most materials exhibit a common three-stage phenomenology, characterized by a decelerating strain-rate during primary creep, possibly followed by a secondary creep with a constant minimum strain-rate (in many materials, this can simply resume to an inflexion point in the strain-rate evolution) that precedes an acceleration of deformation (tertiary creep) leading to macroscopic failure in brittle or quasi-brittle materials, or fluidization in soft matter \citep{siebenburger2012creep} at a creep failure time $t_f$. One overarching goal is therefore to predict, to some extent, this failure time, or lifetime, as a function of the considered material, the applied load, as well as environmental conditions, including temperature. 

To explain such time-dependent deformation and failure under a constant applied load, thermally activated processes have been discussed for a long time (e.g. Ref. \citep{cottrell1952time}). In brittle rocks, subcritical crack or damage growth, which itself depends on temperature and environmental conditions (humidity, presence of other corrosion species...) is considered as a main ingredient of the so-called brittle creep \citep{scholz1968,atkinson1982,atkinson1984,brantut2013}. Consequently, it seems natural to express, at least empirically, the creep lifetime~$t_f$ in the form of an Arrhenius relation \citep{scholz1968,zhurkov1965,pomeau2002,guarino2002,brantut2013}: 
\begin{eqnarray}\label{eqintro001}
t_f \sim t_0 \exp{\left[ \frac{E}{k_BT}   \right]},
\end{eqnarray}
where $E$ is the activation energy, e.g. corresponding to some corrosion reaction in rocks \citep{scholz1968,atkinson1984}, $t_0 = \Omega_0^{-1} \approx 10^{-13} \; \si{\second}$ is the reciprocal of an attempt frequency whose value is of the same order of magnitude for different solids and independent of the structure and chemical nature of the solid \citep{zhurkov1965},  $k_B$ is the Boltzmann constant and $T$ the temperature. In order to make predictions of the creep lifetime, one difficulty is therefore to estimate the appropriate value of the activation energy $E$, and/or to identify a thermally activated microscopic mechanism dominating the creep process and associated to an activation volume $V_a$. A natural way to link the activation volume and the energy is $E=V_a\Delta\sigma$, where $\Delta\sigma=\sigma_{ath}-\sigma$ is the stress gap between a local athermal stress threshold $\sigma_{ath}$, or "strength", and the local stress state $\sigma$, such that the microscopic deformation and damage mechanism takes place athermally for $\sigma=\sigma_{ath}$, when the energy barrier vanishes \citep{castellanos2018,weiss2023logarithmic}. Note, however, that a non linear model, $E \sim \Delta\sigma^{3/2}$, has been proposed as well in case of molecular systems \citep{maloney2006barrier}. 

However, it has been shown that additional difficulties arise when considering material heterogeneity. In this case, some authors argued, on the basis of democratic fiber-bundle models (FBM), that Eq.~\ref{eqintro001} would be valid only if introducing a heterogeneity-dependent effective temperature $T_{eff}$, instead of the thermodynamic temperature $T$, in the Arrhenius expression \citep{ciliberto2001,roux2000,pride2002}. In other words, the effective temperature of the system is an amplification of the thermodynamic one, due to the initial (quenched) spatial disorder of the fiber bundle. Ciliberto~et~al.~\citep{ciliberto2001} concluded that thermal noise and heterogeneity respectively create and enhance the cracking process. Consequently, the larger the spatial heterogeneity of the material, the smaller are the failure times on average (as $T_{eff}$ increases). In addition, for large initial heterogeneities, the failure time is essentially dictated by the quenched spatial heterogeneity, and less sensitive to the thermal noise. 

Nevertheless, these conclusions were obtained from a very simple model of creep rupture, based on some rough assumptions, in particular: (i) a democratic, i.e. mean-field elastic stress redistribution and (ii) an "all or none" breaking rule at the local scale, i.e. absence of memory at the scale of the fiber. In what follows, we explore further the concept of a heterogeneity-dependent effective temperature for the creep of materials, and particularly creep lifetimes, for more realistic situations involving memory effects at the local scale and realistic non-democratic, non-convex elastic redistribution kernels. This allows to analyze the interplay between thermal activation, material heterogeneity, damage mechanics, and the nature of elastic interactions on creep rupture and lifetimes. 

\HL{The approach developed below is particularly suited to study the creep of brittle or quasi-brittle materials, such as rocks or concrete, for which the concept of damage, i.e.~a degradation of local elastic properties during creep, is fully relevant. However, the phenomenology of creep is shared with amorphous media, metallic glasses \citep{zhang2024glasses}, or glassy suspensions \citep{ballesta2016aging}. Strong similarities, such as the interplay between thermal activation, elastic stress redistribution, and material heterogeneity \citep{popovic2021thermally,popovic2022amorphous}, but also important differences, exist between the two categories of materials. First, the creep of amorphous media is generally studied from elastoplastic models in which  localized plastic strains are (thermally or athermally) activated, while the elastic constants remain unchanged \citep{popovic2021thermally,popovic2022amorphous}. This differs from our case, where local damage occurs but no plastic strain is accumulated. We note, however, that both mechanisms lead to similar stress redistribution \citep{eshelby1957determination,eshelby1959elastic}. Consequently, we would expect that the main conclusions of this work may be relevant to a larger class of systems, beyond brittle or quasi-brittle materials. In addition, in our framework, damage only accumulates through creeping time and we do not consider structural relaxation or rejuvenation mechanisms, which potentially play an important role in the case of glasses \citep{zhang2022glasses,xu2024stochastic}.}

In the first part (Section \ref{chap_analyt}), we propose a theoretical analysis that shows, in the case of an heterogeneous material while neglecting elastic interactions, that the creep lifetime follows an Arrhenius expression (Eq.~\ref{eqintro001}) with an effective temperature increasing with the heterogeneity \HL{as found by Refs. }\citep{roux2000,ciliberto2001}\HL{.} To introduce realistic elastic interactions and local memory effects, we later (Section \ref{chap_num}) perform a numerical study based on a \HL{two-dimensional} progressive damage model developed by Refs.~\citep{amitrano1999, amitrano2006, girard2010, girard2012} in which we implemented thermal activation~\citep{weiss2023logarithmic,makinen2023history} using a kinetic Monte Carlo algorithm~\citep{bortz1975, kratzer2009}. \HL{The mechanism of the numerical model is damage, and we do not take into account plastic deformation or fracture mechanics as the model is always continuous}. This confirms the validity of the concept of an effective temperature $T_{eff}$, which however depends both on material heterogeneity and the nature of the elastic interaction kernel. Although this damage model and our results are mainly discussed in the context of brittle creep \citep{scholz1968,brantut2013} of materials such as rocks, we believe that our main conclusions would remain valid for other materials, as long as material heterogeneity, thermal activation, and elastic stress redistribution interplay.

In order to avoid confusion, in what follows, material \textit{heterogeneity} refers to spatial variations of the local strength $\sigma_{ath}$, while \textit{disorder} represents a measurement of the entropy of a system ('thermodynamic' disorder) or the number of possible microstates in that system~\citep{pathria2011, sethna2021entropy}. Besides, $\langle X \rangle = N^{-1} \sum_k^{N} (X_k Y_k)$ refers to an arithmetic mean value weighted by a given density distribution $Y_k$ over the $N$ elements (spatial average, at some given time), whereas $\overline{X} = \mathcal{N}_f^{-1} \sum_k^{\mathcal{N}_f} X_k $ refers to an arithmetic mean value over the $\mathcal{N}_f$  transitions needed to reach the macroscopic rupture (time average over the creep process).

\begin{figure}[bt!]
\centering
\includegraphics[width=1.15\columnwidth,center]{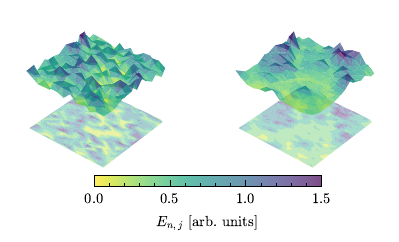}
\caption{\label{fig:entropy00} Representation of the energy barrier field for a system composed of $N$ subvolumes, at different transitions $j$. Left: $E_{n,0}$. Right: $E_{n,j}$}
\end{figure}

\section{Theoretical analysis}\label{chap_analyt}

We consider a volume of a solid submitted to a constant applied stress, which damages and deforms under the combined action of mechanical loading and thermal activation, i.e. experiences creep deformation. Such a volume can be divided into $N$ sub-volumes, each $n=1, ..., N$ micro-element with energy \HL{barrier} $E_{n,j}$ having a probability $p(E_{n,j})$ to be thermally activated and which may damage several times, this activation defining the $(j+1)$th transition by thermal activation for the whole volume. 

\subsection{Rupture time analysis}\label{chap_analyt_tf_Th}

The rupture of the whole sample takes place by the accumulation of successive local damage/microfracturing events. Each event $j+1$ with $j\in[0,\mathcal{N}_f - 1]$ represents a transition of the entire system due to a thermal activation. The average lifetime of a single transition is $\langle \Delta t \rangle_{j+1}$, from which the creep lifetime $t_f$ of the full volume can be obtained, $t_f = \sum_{j=0}^{\mathcal{N}_f-1} \langle \Delta t \rangle_{j+1} $, where $\mathcal{N}_f$ is the number of successive thermally activated events leading to system-size failure\HL{, which happens during tertiary creep, when the strain rate diverges}. 
In this framework, the jump rate $\omega_{j+1}$ associated to event $j+1$ follows an Arrhenius expression \citep{vineyard1957, kratzer2009}:
\begin{eqnarray}\label{eqtheor000a}
\omega_{j+1} \approx \Omega_0 \exp{\left(-\frac{F_{j}}{k_B T} \right)},
\end{eqnarray}
where the subindex $_{j+1}$ represents the $(j+1)th$ transition by thermal activation within the sample, $F_{j}$ is the free energy barrier, inherited from the previous transition, which must be overcome (by thermal activation) to trigger this event, and $\Omega_0$ is the thermal vibration frequency.
The inversion of Eq.~\ref{eqtheor000a} gives the average waiting time between successive transitions, $\langle \Delta t \rangle_{j+1} = \omega_{j+1}^{-1}$~\citep{vineyard1957}: 
\begin{eqnarray}\label{eqtheor000c}
\langle \Delta t \rangle_{j+1} = \frac{1}{\Omega_0} \exp{\left(\frac{F_{j}}{k_B T} \right)}.
\end{eqnarray}

Here the waiting time, $\Delta t_{j+1}$, between transitions follows an exponential distribution \citep{fichthorn1991, walpole2007}, interpreted in the study of solids as a transitional probability of damage/fracturing under a constant load \citep{cruden1970},
\begin{eqnarray}\label{eqtheor000d}
f(\Delta t_{j+1})_{j+1} = \omega_{j+1} \exp{\left(-\omega_{j+1} \Delta t_{j+1} \right)},
\end{eqnarray}
 with a mean and standard deviation equal to $\omega_{j+1}^{-1}$. It is worth stressing here that after each event, the jump rate $\omega_{j+1}$ is modified, as a damage event redistributes elastic stresses in the rest of the volume, modifying the energy landscape to $E_{n,j+1}$, and therefore the waiting time to the next transition. This coupling between thermal activation and mechanical interactions between sub-volumes and events is a major topic of the present work.  Then, the average rupture time and its variance, $\langle t_f \rangle$ and $\text{Var}[t_f]$, are given by:
\begin{eqnarray}\label{eqtheor000e}
\langle t_f \rangle = \sum_{j=0}^{\mathcal{N}_f-1} \langle \Delta t \rangle_{j+1} = \frac{1}{\Omega_0} \sum_{j=0}^{\mathcal{N}_f-1} \exp{\left(\frac{F_{j}}{k_B T} \right)}, \\
\text{Var}[t_f] = \sum_{j=0,k=0}^{\mathcal{N}_f-1} \text{Cov}[\Delta t_{j+1}, \Delta t_{k+1}],
\end{eqnarray}
i.e., the average and the variance of a sum of random variables \citep{walpole2007}. Note that $\text{Var}[t_f] = \sum_{j=0}^{\mathcal{N}_f-1} \langle \Delta t \rangle_{j+1}^2$ for independent transitions \citep{walpole2007}. In addition, the probability distribution of the failure time $f(t_f)$ could be represented as the convolution operation between all the $f(\Delta t_{j+1})$~\citep{sornette2006},  
which converges towards a Gaussian density distribution for a very large number of independent transitions, $j \to \infty$, following the central limit theorem~\citep{sornette2006}. However, in the considered systems, a given transition $(j+1)$ is dependent of what occurred at all the previous $j$ transitions. In other words, the transitions are not \textit{a priori} independent, and therefore a non-Gaussian density distribution of failure times could potentially emerge as shown by Ref. \citep{alava2021timedistribution} in the case of Weibull distributed intial lifetimes.

\subsection{Spatial heterogeneity analysis}\label{chap_analyt_thermo}

Spatial heterogeneity of the material is introduced at the scale of the $N$ subvolumes (indexed by $n$), each one associated to a variable activation energy $E_{n,j}$ or energetic state, which is also evolving with the transition number $j$ as the result of previous transitions and their associated stress redistribution (see Fig.~\ref{fig:entropy00}).

As we consider a mechanical system, we link the activation energy $E_{n,j}$ to a local stress gap $\Delta\sigma_{n,j}$, i.e. $E_{n,j}= \Delta\sigma_{n,j} V_a$, where $V_a$ is the activation volume for the considered damage/fracturing microscopic mechanism, considered here as being constant in space and time. The stress gap $\Delta\sigma_{n,j}$ represents a distance, \HL{in a similar way as in Bouchaud's trap model} \cite{bouchaud1992disordered}\HL{,} expressed in the principal stress space, between the local threshold $\sigma_{ath,n}$ (hereafter denoted athermal as it corresponds to a failure threshold in absence of thermal disorder) and the local stress state, $\Delta\sigma_{n,j} \sim \sigma_{ath,n} - \sigma_{n,j}$. Here we do not address how the stress on each element may vary in space, and this democratic hypothesis leads to consider that stress is initially equal in each element. Material heterogeneity is introduced at the level of each subvolume $n$ through $\sigma_{ath,n}$ considered as a quenched random variable~\citep{makinen2023history, weiss2023logarithmic}. \HL{The linear dependence of the energy barrier $E_{n,j}$ on the stress gap $\Delta\sigma_{n,j}$ is equivalent to the Eyring's model \citep{eyring1935activated}. It has been argued, however, that a non-linear dependence, $E_{n,j} \sim \Delta\sigma_{n,j}^{3/2}$, would be more appropriate in case of driven molecular systems such as glasses \citep{maloney2006barrier}. We considered a linear model in most of what follows. However, in the numerical analysis, we compared simulations performed with either a linear or a non-linear model, and found that our main conclusions are independent of this scaling between the activation energy and the stress gap (see section \ref{chap_num}).
}

From the definition of the average waiting time between transitions, $\langle \Delta t \rangle_{j+1} \sim \exp{(F_{j}/k_BT)}$, the free energy of the system is defined as ~\citep{sornette2006, pathria2011, sethna2021entropy}:, 
\begin{eqnarray}\label{eqtheor001e}
F_j = -k_BT\log{Z_j}.
\end{eqnarray}

where $Z_j = \sum_n e^{-E_{n,j}/k_B T}$ is the partition function~\citep{sornette2006, pathria2011, sethna2021entropy}.
As an alternative representation of the last equations, it is more convenient to introduce the density of states $g(E_{n_b,j})$, which corresponds to the number of energetic states, i.e. the number of the sub-volumes, that have an energy within a small range $E_{n_b,j} \pm dE_{n_b,j}$, with $n_b$ representing the number of the bin in which the energy value $E_{n_b,j}$ falls  \citep{sornette2006, pathria2011}. We can then rewrite for a given bin $n_b$, $g(E_{n_b,j}) dE_{n_b,j} = N f(E_{n_b,j}) dE_{n_b,j}$, which can be replaced into the partition function \citep{pathria2011}:
\begin{eqnarray}\label{eqtheor001c}
Z_j = N \sum_{n_b} f(E_{n_b,j}) e^{-E_{n_b,j}/k_B T},
\end{eqnarray}
where $f(E_{n_b,j}) = g(E_{n_b,j}) / N \approx f(\Delta\sigma_{n_b,j}) / V_a$ represents a relative density of states and is an explicit representation of the spatial heterogeneity in the system. It is equivalent to the probability distribution of the stress gap (also called the excitation spectrum), taking into account that the activation volume is assumed constant here.

So far, we did not make any assumption about the nature of the heterogeneity. To go further, we assume below a Gaussian distribution of local stress gaps $\Delta \sigma_j$ to obtain a prediction of the failure time (Section \ref{chap_analyt_tf02}). The case of a uniform distribution is detailed in Appendix \ref{app:b}, giving very similar results.

\begin{figure}[h]
\centering
\includegraphics[width=\columnwidth]{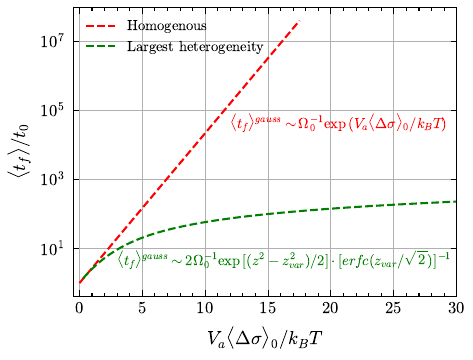}
\caption{\label{fig:theor2} Dependency of the normalized average lifetime $\langle t_f \rangle / t_0$ on $V_a\langle\Delta\sigma\rangle_0/k_BT$, for a Gaussian initial distribution of stress gaps, and mechanical interactions neglected. Theoretical result.}
\end{figure}

\subsection{Prediction of the rupture time $t_{f}$ and the effective temperature $T_{eff}$ for a heterogeneous system}\label{chap_analyt_tf}

We switch to the main question of this work: What is the combined effect of material heterogeneity (variability of the stress gaps $\Delta \sigma_j)$ and (thermal) disorder  on creep lifetimes? And, do we still get an Arrhenius-like expression for $t_f$, possibly with an heterogeneity-dependent effective temperature $T_{eff}$?

Replacing Eq. \ref{eqtheor001c} in Eq. \ref{eqtheor001e}, and later replacing it in Eq. \ref{eqtheor000e}, we can identify an Arrhenius-like expression for the average failure time $\langle t_f \rangle = t_0 \exp{(E_{\Delta \sigma_{h,0}}/k_BT_{eff})}$, with the effective temperature $T_{eff}$ and the "characteristic" timescale $t_0$ of Eq.~\ref{eqintro001} given respectively by (details in appendix \ref{chap_analytApp_tf}):
\begin{eqnarray}\label{eqtf003}
k_B T_{eff} = \frac{k_B T}{ 1 - \Delta F_{cor} / E_{\Delta \sigma_{h,0}}}, \\
t_0 = \frac{1}{\Omega_0}\frac{\mathcal{N}_{f}}{N},
\end{eqnarray}

where $E_{\Delta \sigma_{h,0}}=\langle E_{\Delta\sigma}\rangle_{0}=V_a \langle\Delta\sigma\rangle_{0}$ is the initial arithmetic mean energy barrier and the term $\Delta F_{cor}$ corresponds to a corrected differential free energy given by: 
\begin{multline}\label{eqtf004}
\Delta F_{cor} = - k_BT \log{ \left[  \overline{\exp{\left(-\frac{\Delta F_{j}}{k_B T} \right)}} \right]} \\
= E_{\Delta \sigma_{h,0}} - k_BT \log{ \left[  \overline{\frac{N}{Z_j}} \right]},
\end{multline}
and $\Delta F_{j} = F_{h,0} - F_{j} \geq 0$ is the differential free energy between the "equivalent" homogeneous free energy at the beginning of the experiment, $F_{h,0} = E_{\Delta \sigma_{h,0}} - k_BT\log{N}$, and the free energy after a transition $j$ is triggered by thermal activation. $\Delta F_{j}$ is a measure of the heterogeneity of a system, just before its transition $j$, and two extreme cases appears: the first one occurs when $\Delta F_{j} = 0$ $\forall j$, which would indicate that the system remains homogeneous during the entire process, i.e. $\Delta F_{cor} = 0$. The other one \HL{is given by the fact that the effective temperature should be positive}, i.e. $T_{eff} > 0$, which indicates that $\Delta F_{cor} < E_{\Delta \sigma_{h,0}}$. Hence, the correcting term, $\Delta F_{cor}$, varies between two limits, i.e. $E_{\Delta \sigma_{h,0}} > \Delta F_{cor} \geq 0$. 

Eq. \ref{eqtf003} indicates that, even in a very general case, an effective temperature could be defined, entering an Arrhenius-like expression for the creep failure time. However, the term $\Delta F_{cor}$ is complex and depends on the entire creep process, and particularly the evolution of the stress gap distribution, as individual sub-volumes are, thermally or athermally, activated, leading to elastic stress redistributions. In other words, at this stage, a simple expression between $T_{eff}$ and the characteristics of material heterogeneity cannot yet be proposed. Nevertheless, as found by Ref.~\citep{ciliberto2001}, here it is possible to indicate that for an invariant homogeneous system $k_B T_{eff} = k_B T$, while, on the reverse, the effective temperature increases as the heterogeneity increases. 

We can obtain a new expression for the failure time, $\langle t_f \rangle$, as $\langle t_f \rangle = t_0 \exp{(E_{\Delta \sigma_{h,0}}/k_BT)} \exp{(-\Delta F_{cor}/k_BT)}$, where the term $e^{-\Delta F_{cor}/k_BT}$, can be interpreted as a correcting factor around an initial homogeneous "equivalent" system. 
Additionally, the upper and lower limits of the creep lifetimes are given by $t_0 \exp{(E_{\Delta \sigma_{h,0}}/k_BT)} \geq \langle t_f \rangle > t_0$.
For an initially homogeneous system, \HL{in the course of creep deformation and as the result of damage and elastic interactions}, this term takes a value close to but different from 1. \HL{Indeed, if we take into account that just one thermally activated damage event implies a stress redistribution to the surrounding elements, the stress field and therefore the stress gap field become progressively heterogeneous.}

\begin{figure}[t!]
\centering
\includegraphics[width=\columnwidth]{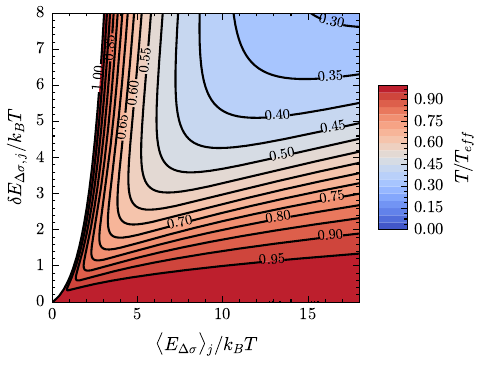}
\caption{\label{fig:theor3} Example of the effective temperature term $T/T_{eff}$ as a function of the arithmetic mean value and the standard deviation of the initial distribution of the stress gap.}
\end{figure}

\subsubsection{Negligible mechanical interactions: general case}\label{chap_analyt_tf01}

In order to express the effective temperature $T_{eff}$ and the rupture time $t_{f}$ as a function of the initial material heterogeneity for the most simple case, we first make a strong assumption, considering that all the transitions are independent, i.e.~
the probability distribution of the stress gaps, $f(\Delta\sigma_{n_b,j}) = f(\Delta\sigma_{n_b,0})$, does not evolve during creep. In particular, it is not affected by thermal activation, and elastic stress redistributions are neglected. In other words, there is no memory of the past events and no damage localisation. In this case, creep becomes a Poisson process with a rate $\omega_j = \omega_1$ and the probability density of waiting times between successive events is exponential \citep{fichthorn1991, walpole2007}. Then, the failure will be defined by a maximum number of thermally activated events $\mathcal{N}_f$ accumulated during the experiment.

With such a (rough) assumption, we can write an expression for the predicted average failure time $\langle t_{f} \rangle$ and its variance $\text{Var}[t_f]$ as a function of the initial material heterogeneity as:
\begin{eqnarray}\label{eqtf006}
\langle t_{f} \rangle \approx \mathcal{N}_{f} \langle \Delta t \rangle_1 \approx \frac{\mathcal{N}_{f}}{\Omega_0}  \exp{\left( \frac{F_0}{k_B T} \right)}, \\
\text{Var}[t_f] \approx \sum_{j=1}^{\mathcal{N}_f} \langle \Delta t \rangle_1^2 \approx \mathcal{N}_{f} \langle \Delta t \rangle_1^2,
\end{eqnarray}
where  $F_0$ corresponds to the initial free energy, and \HL{$ e^{F_{0}/k_BT} = 1 / [N \sum_{n_b} f(E_{n_b,0}) e^{-E_{n_b,0}/k_B T}]$}, which determines the value of the average duration of the first time step $\langle \Delta t \rangle_1$. We can see additionally that the average creep lifetime and its standard deviation are linearly related as $\delta t_f = \sqrt{\text{Var}[t_f]} = \langle t_{f} \rangle / \sqrt{\mathcal{N}_f}$. 

\subsubsection{Negligible mechanical interactions: Gaussian quenched heterogeneity}\label{chap_analyt_tf02}

To be slightly more quantitative, and taking into account that the material heterogeneity is introduced trough the athermal threshold $\sigma_{ath,n}$, we could say that the density distribution of the stress gap $f(\Delta\sigma_j)$ is of the same shape as the one of the athermal threshold $f(\sigma_{ath,j})$ as we do not consider elastic redistribution in the theoretical analysis, i.e.~the stress $\sigma_{n,j}$ equals everywhere.

Consequently, we now consider the case of a quenched Gaussian distribution of the stress gaps, i.e. Gaussian distribution of the energy barrier $E_j\sim\mathcal{N}(\langle E_{\Delta\sigma}\rangle_j,\delta E_{\Delta\sigma,j}^2)$ with arithmetic mean value and variance $\langle E_{\Delta\sigma} \rangle_j$, $\delta E_{\Delta\sigma,j}^2$.
Then, the probability density $f(E_j)$ reads:
\begin{eqnarray}\label{eqtheor001l}
f(E_j) = \frac{1}{\delta E_{\Delta\sigma,j}\sqrt{2\pi}} \exp{\left[ -\frac{1}{2} \left( \frac{E_j - \langle E_{\Delta\sigma} \rangle_j}{\delta E_{\Delta\sigma,j}}  \right)^2 \right]}.
\end{eqnarray}

We can then obtain an expression for the lifetime if we replace the obtained expression for the free energy in the case of a Gaussian probability density of stress gap, $F_0/k_B T =  \langle E_{\Delta\sigma} \rangle_0/k_B T - 0.5 \left( \delta E_{\Delta\sigma,0}/k_B T \right)^2 - \log{\left[  0.5N \erfc{\left( z_{var,0}/\sqrt{2} \right)} \right]}$, into Eq.~\ref{eqtf006} (solution in Appendix \ref{app:a}):
\begin{multline}\label{eqtheor005}
\langle t_{f} \rangle^{gauss} = \frac{2}{\Omega_0} \frac{\mathcal{N}_{f}}{N} \exp{\left(\frac{\langle E_{\Delta\sigma} \rangle_0}{k_BT}\right)} \times \\
\exp{\left[ -\frac{1}{2} \left( \frac{\delta E_{\Delta\sigma,0}}{k_BT} \right) ^2 \right]}   \left[\erfc{\left( \frac{z_{var,0}}{\sqrt{2}} \right)}  \right] ^ {-1},
\end{multline}
where $z_{var,0} =  \delta E_{\Delta\sigma,0} / k_B T  -  \langle E_{\Delta\sigma} \rangle_0 /\delta E_{\Delta\sigma,0}$ and $\erfc{(\textbf{x}^*)}=2/\sqrt{\pi}\int_{\textbf{x}^*}^{\infty} e^{-x^2}  dx$, the Gauss complementary error function. This shows that the heterogeneity plays an important role on creep lifetime. We can also deduce from this expression that for a very high temperature
, all the elements are activated at the same time from the beginning, then $\langle t_{f}\rangle^{gauss} \sim \Omega_0^{-1}$. As expected, in case of a homogeneous system ($\delta_{\Delta\sigma,0} \to 0$), the failure time follows a "classical" Arrhenius expression, with the thermodynamic temperature $T$, $\langle t_{f} \rangle^{gauss} \sim \Omega_0^{-1} \exp{\left(\langle E_{\Delta\sigma} \rangle_0 / k_B T \right)}$\HL{.} On the reverse, for a strong heterogeneity, $\langle t_{f}\rangle^{gauss} \sim 2 \Omega_0^{-1}\exp{[(z^2 - z_{var,0}^2) / 2]} [\erfc{(z_{var,0}/\sqrt{2})}]^{-1}$ (taking into account ~$\delta_{\Delta\sigma,0} \gg 0$ and $\Delta\sigma \geq 0$, i.e. $\langle\Delta\sigma\rangle_0 \approx z\delta_{\Delta\sigma,0}$ with $z$ the $z-$value of the confidence interval. $z=1.96$ for $95\%$ of confidence interval). These two expressions represent upper and lower bounds for creep lifetimes for a Gaussian quenched heterogeneity, and memory effects neglected. They are equivalent to those found by Ref.~\citep{ciliberto2001} (see Fig.~\ref{fig:theor2}).

Doing an analysis similar to that of Section \ref{chap_analyt_tf01}, we find that the effective temperature for a system with a Gaussian quenched heterogeneity follows Eq.~\ref{eqtf003} with $\Delta F_{cor}$ expressed by (see Fig.~\ref{fig:theor3}):
\begin{equation}\label{eqtheor006}
\frac{\Delta F_{cor}}{k_B T} = \frac{\Delta F_0}{k_B T} \approx \frac{1}{2} \left(\frac{\delta E_{\Delta\sigma,0}}{k_BT}\right )^2 + \log{ \left[ \frac{1}{2} \erfc{\left( \frac{z_{var,0}}{\sqrt{2}} \right)} \right]}.
\end{equation}

\begin{figure}[t!]
\centering
\includegraphics[width=\columnwidth]{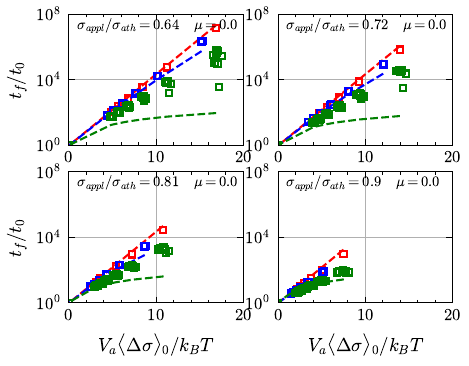}
\caption{Theoretical (dashed line) vs Numerical (squares) results for the normalized lifetime $t_f / t_0$  as a function of the normalized arithmetic mean value of the energy barrier $\langle E_{\Delta\sigma} \rangle_0 / k_BT$. Friction coefficient $\mu = 0$ and different values of the stress ratio $\sigma_{appl} / \sigma_{ath}$. Red: $\delta C = 0$, blue: $\delta C = 1$, green: $\delta C = 5$}
\label{fig:num02}
\end{figure}

\section{Numerical analysis}\label{chap_num}

\subsection{Numerical model}\label{sect_numMod}

What we learned from the previous sections can be summarized as follows:
In the most general case, our analysis still suggests an Arrhenius-like expression, however with an effective temperature $T_{eff}$ explicitly increasing with the heterogeneity and depending on both the thermodynamic temperature and, in a non-trivial way, the entire evolution of the excitation spectrum during the creep process. This evolution results from the modification of the internal stress field after each thermally activated event, and possible athermal cascades of events (avalanches). 

To see if a generic and simple dependence of $T_{eff}$ on heterogeneity still holds in this complex situation, we rely below on numerical simulations based on a progressive damage model ~\citep{amitrano1999, amitrano2006, girard2010, girard2012} with thermal activation \citep{makinen2023history, weiss2023logarithmic}. This model \HL{is not scalar and} accurately introduces realistic elastic interaction kernels from finite element analysis, memory effects through a damage parameter, and thermal activation from a Bortz--Kalos--Lebowitz (BKL) kinetic Monte Carlo (KMC) algorithm \cite{bortz1975}. Compared to the much simpler democratic fiber bundle model analyzed by Ciliberto~et~al.~\cite{ciliberto2001} and Roux \cite{roux2000}, our model differs in different ways, including a non-mean-field (non-democratic) and non-convex elastic redistribution kernel \cite{dansereau2019collective}, and the fact that a given sub-volume is never fully broken, i.e. it can damage several times during the creep process. The goal of the KMC algorithm is to thermally activate damage events at the element scale according to the rates of the relevant individual processes. The jump rate of each element $n$ follows an Arrhenius expression~\citep{kratzer2009, bortz1975} $\omega_{n, j+1} = \Omega_0\exp\left(-E_{n,j}/k_BT\right)$,
where $E_{n,j} = V_a \Delta\sigma_{n,j}$ and $\Omega_0$ is a vibration frequency, which is, in solids, of the order of $10^{13}\; \si{\second}^{-1}$ \citep{zhurkov1965} and $V_a$ is a time- and spatially-invariant activation volume. In the present case, in order to simulate brittle creep \citep{scholz1968}, we used a Coulomb stress gap between a local cohesion $C_n$ and the local Coulomb stress, $\Delta\sigma_{n,j}\sim C_n-(\tau_{n,j}-\mu\sigma_{n,j})$ (Fig.~\ref{fig:pdm02}), \HL{where $\tau_{n,j}$ and $\sigma_{n,j}$ represent respectively the shear and the normal stress over an orientation that maximizes the Coulomb stress $\tau_{n,j}-\mu \sigma_{n,j}$, $\phi$ is the friction angle, and $\tan\phi=\mu$ the corresponding internal friction coefficient. After one element thermally activates, the stress and strain fields are recomputed and avalanches occur athermally as long as the local damage (Mohr--Coulomb) criterion is satisfied in one or more elements, $\Delta\sigma_{n,j}<0$. We assume a clear separation of timescales between the short timescale of damage propagation during an avalanche and the much longer activation timescales. Consequently, time is stopped during an avalanche. The simulation stops when the macroscopic axial strain reaches a value of $2\times10^{-2}$, defining the failure time $t_f$ and the total number of transitions~$\mathcal{N}_f$. This occurs during the tertiary creep stage, when the deformation rate diverges, $\dot{\varepsilon}_j \to \infty$.} To take into account microstructural heterogeneity \citep{hansen19904}, the cohesion $C$ is set randomly, and then kept unchanged to mimic quenched material heterogeneity, for each element from a Gaussian distribution, $C_n \sim \mathcal{N}(\langle C \rangle, \delta C^2)$, with arithmetic mean value $\langle C \rangle$ and standard deviation $\delta C$,  $\langle C \rangle$ being constant for all the cases and $\delta C$ an explicit representation of the heterogeneity. \HL{The distribution is truncated to avoid negative cohesion cases} (see appendix \ref{app:c} for additional details about the model).

The results shown below were obtained for a system of dimension $x\times y = 0.5 \times 1$ divided into 960 triangular elements (sections \ref{chap_numres01} and \ref{chap_numres03}) and 3968 triangular elements (sections \ref{chap_numres02}) \citep{girard2010}, but we checked that the main conclusions of this work were size-independent. For a given configuration (internal friction $\mu$, initial heterogeneity $\delta C$), we first run an athermal test to obtain the corresponding strength~$\sigma_{ath}$, and later setting the applied creep stress , $\sigma_{appl}$, at different percentages of the athermal strength. This percentage represents the stress ratio $\sigma_{appl} / \sigma_{ath} < 1$. For the analysis of the effective temperature and the activation volume (in \ref{chap_numres01} and \ref{chap_numres03}) we performed 10 realizations of the microstructure in each specific configuration (stress ratio $\sigma_{appl} / \sigma_{ath}$, temperature $T$, and heterogeneity $\delta C = 0, 1, 5 \; \si{\mega\pascal}$, for both $\mu=0.7$ and $\mu = 0$). Additionally, for the analysis of the probability distribution of the failure time and particularly its variability around its arithmetic mean value (in \ref{chap_numres02}), we performed 20 realizations of the microstructure in each configuration (stress ratio $\sigma_{appl} / \sigma_{ath}$, temperature $T$, heterogeneity $\delta C = 0$ and $5 \; \si{\mega\pascal}$, $\mu=0.7$). "Mirror" creep experiments are then performed, using exactly the same quenched heterogeneity and configuration, but introducing different realizations of thermal activation (20~realizations). For more details about the different input and output parameters of the creep model the with their respective values, see Table \ref{t1} in Appendix~\ref{app:d}.

\begin{figure}[t!]
\centering
\includegraphics[width=\columnwidth]{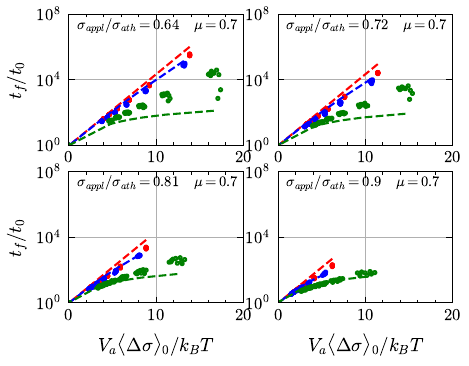}
\caption{Same as Fig. \ref{fig:num02} for $\mu=0.7$.} 
\label{fig:num03}
\end{figure}

\subsection{Analysis of results}\label{sect_num}

\subsubsection{Effective temperature and dependency on the heterogeneity and the elastic kernel}\label{chap_numres01}

In Figs. \ref{fig:num02} and \ref{fig:num03}, we compare our numerical results, expressed as individual lifetimes $t_f$ obtained for each simulation, normalized by the inverse of $t_0$ in a semilog scale as a function of the argument $V_a \langle \Delta \sigma \rangle_0 / k_BT$. The results obtained for $\mu=0$ (squares) and $\mu=0.7$ (dots) and different values of the stress ratio $\sigma_{appl}/\sigma_{ath}$ are compared with the theoretical predictions (dashed lines) obtained previously for a Gaussian heterogeneity while ignoring elastic stress redistribution and memory effects (Eq.~\ref{eqtheor005}).

For a given temperature $T$ and stress ratio, the failure time decreases as the heterogeneity increases in all cases, thus extending this key result, obtained previously for a democratic FBM \citep{ciliberto2001,roux2000} as well as from our theoretical predictions. 
One additional remark is that for high temperatures, the rupture time tends to a constant value, $t_{f} (T \to \infty) \approx t_0 = \mathcal{N}_f /(N\Omega_0)$, i.e., independently from the heterogeneity.

For homogeneous systems, i.e.~$\delta C=0$, the numerical results are close to the theoretical curves, which themselves follow a classical Arrhenius relation ruled by the thermodynamic temperature $T$  (see Section \ref{chap_analyt_tf02}). This means that, in this case, although the system becomes progressively heterogeneous in terms of local elastic properties as the result of damage, this mechanism has a limited (however not completely negligible, see more below) effect on creep lifetimes. On the other hand, for systems with strong heterogeneity, the analytical prediction underestimates the real value of $t_f$ obtained from the numerical simulations, and this discrepancy increases upon decreasing the stress ratio $\sigma_{appl}/\sigma_{ath}$, meaning that the stress redistribution and memory effects plays an important role in creep lifetimes. 

If one assembles in a single plot all the data of Figs.~\ref{fig:num02} and \ref{fig:num03}, we obtain highly scattered values. This confirms that creep lifetimes of heterogeneous brittle materials do not follow a classical Arrhenius relation with the \textit{thermodynamic} temperature $T$ as a controlling parameter. The question is therefore: Can a generic Arrhenius-like relation, with an heterogeneity-dependent effective temperature $T_{eff}$ substituting for $T$, unify the results obtained for different degrees of heterogeneity and internal frictions?, i.e.: 
\begin{equation}\label{eqres003}
t_f \approx \frac{1}{\Omega_0} \frac{\mathcal{N}_{f}}{N} \exp{\left( \frac{V_a \langle \Delta \sigma \rangle_0}{k_B T_{eff}} \right)}
.
\end{equation} 

\begin{figure}[!]
    \begin{subfigure}[b]{\columnwidth}
        \centering
        \includegraphics[width=\columnwidth]{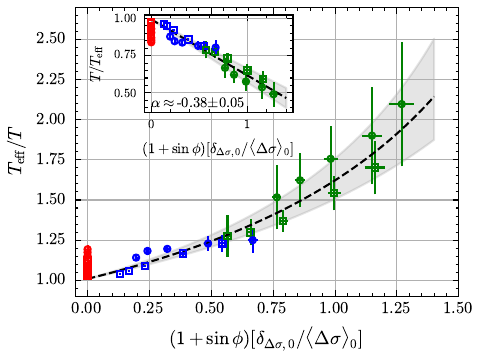}
        \caption{\label{fig:num04a} $T_{eff} / T = f(\delta_{\Delta\sigma,0} / \langle \Delta\sigma \rangle)$.}
    \end{subfigure}
    \vfill
    \begin{subfigure}[b]{\columnwidth}
        \centering
        \includegraphics[width=\columnwidth]{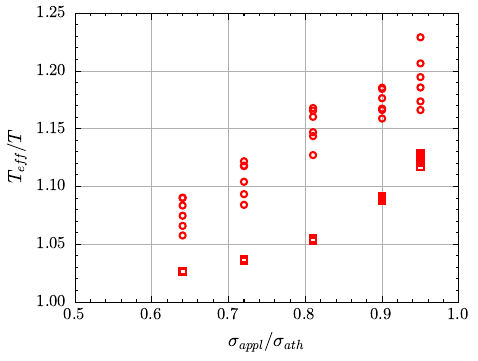}
        \caption{\label{fig:num04b} $T_{eff}^{homog} / T = f(\sigma_{appl}/\sigma_{ath})$}
    \end{subfigure}
    \caption{\HL{a) Effective temperature ratio $T_{eff} / T$ as a function of the heterogeneity. Red: $\delta C = 0$, blue: $\delta C = 1$, green: $\delta C = 5$. b) Effective temperature ratio $T_{eff}/T$ as a function of the stress ratio $\sigma_{appl}/\sigma_{ath}$ for homogeneous systems. Squares: $\mu = 0.0$, dots: $\mu = 0.7$. }}
    \label{fig:num04}
\end{figure}

\begin{figure}[t!]
\centering
\includegraphics[width=\columnwidth]{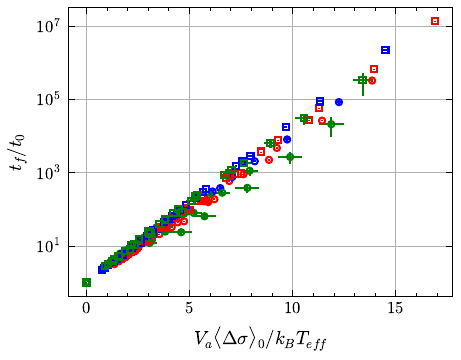}
\caption{Dependency of the lifetime $t_f / t_0$ on $V_a\langle\Delta\sigma\rangle_0/k_BT_{eff}$. Squares: $\mu = 0.0$, dots: $\mu = 0.7$. Red: $\delta C = 0$, blue: $\delta C = 1$, green: $\delta C = 5$.}
\label{fig:num04c}
\end{figure}

Figs. \ref{fig:num04} and \ref{fig:num04c} positively answers to this conjecture. Fig.~\ref{fig:num04a} shows that an empirical but generic relationship between the temperature enhancement factor $T_{eff}/T$ (obtained from Eq.~\ref{eqtf003}) in one hand, the quenched heterogeneity quantified by \HL{$\delta E_{\Delta \sigma,0}/\langle E_{\Delta \sigma} \rangle_0 \propto \delta_{\Delta\sigma,0}/\langle \Delta \sigma \rangle_0$} as well as the internal friction angle $\phi$ in the other hand, can be proposed. It reads:
\begin{equation}\label{eqres001}
k_BT_{eff} \approx \frac{k_BT}{1 - \alpha (1+\sin\phi) [\delta_{\Delta\sigma,0}/\langle \Delta \sigma \rangle_0]},
\end{equation}
where $\alpha = 0.38 \pm 0.05$ is obtained from a linear regression. The error bars correspond to the variability of $T_{eff}/T$ for the same conditions of temperature, load and internal friction but different realizations of quenched heterogeneity. 
The form of this relation is similar to that proposed by Ciliberto et al. \citep{ciliberto2001} for a democratic FBM (their Eq.~16), however with an important difference: in our case, the effective temperature depends not only on the quenched heterogeneity but also on the internal friction, i.e. on the nature of the elastic redistribution kernel. \HL{We performed the same analysis from a set of simulations using a non-linear energy barrier scaling, $E_{n,j}\sim\Delta\sigma_{n,j}^{3/2}$ \cite{maloney2006barrier}, obtaining a similar expression but with $\alpha = 0.61 \pm 0.15 \approx 3/2 \times 0.38$. 
The factor 3/2 comes from the approximate relation
$\delta_{\Delta \sigma^{3/2}}/\langle \Delta \sigma^{3/2} \rangle \approx (3/2) \delta_{\Delta \sigma}/\langle \Delta \sigma \rangle$.
The form of Eq. \ref{eqres001} is then independent of the energy barrier model, except for a change in the parameter~$\alpha$. A non-linear energy barrier model reinforces the heterogeneity effect on the effective temperature, i.e.~$\alpha$ increases.}

Several conclusions can be drawn from Eq.~\ref{eqres001}:

(i) This expression predicts that $T_{eff} \rightarrow T$ when $\delta_{\Delta\sigma,0}/\langle \Delta \sigma \rangle_0 \rightarrow 0$, in agreement with our former analytical derivations (see Eq.~\ref{eqtf003} and Section \ref{chap_analyt_tf01}) and previous work on democratic FBM \citep{ciliberto2001,roux2000} for homogeneous systems. However, we see from Fig.~\ref{fig:num04a} that this is only approximately true. More precisely, for purely homogeneous systems (in terms of cohesion values), the ratio $T_{eff} / T$ actually ranges between 1 and 1.25, i.e.~there is a limited but significant effect of damage on effective temperature enhancement. This effect is more pronounced for $\mu=0.7$ and for stress ratio $\sigma_{appl}/\sigma_{ath}$ close to 1, i.e.~when the applied creep load is near the athermal strength (Fig.~\ref{fig:num04b}). 

(ii) In agreement with our theoretical expectation (see Section \ref{chap_analyt_tf}) as well as previous results obtained from a democratic FBM \citep{ciliberto2001,roux2000}, the effective temperature increases with increasing quenched heterogeneity, and this dependence sums up to an empirical but generic and simple relationship. This therefore extends this concept to more complex situations involving more realistic, non-democratic elastic redistribution kernels and memory effects.

(iii) The effect of material heterogeneity on $T_{eff}$ is actually coupled to an effect of the elastic kernel. The role of quenched heterogeneity is not modified for a plastic-like (Tresca) kernel, i.e.~$\mu=0$ ($\sin\phi=0$), but is reinforced for increasing values of $\mu$. We can argue that increasing $\mu$ modifies the geometry of the quadrupolar (non-convex) elastic redistribution kernel~\citep{amitrano1999}, meaning that, after a damage event, elastic stresses are redistributed along narrower but more extended branches, i.e.~the fractal dimension of the damage field decreases~\citep{amitrano1999}. This reinforces the microstructural heterogeneity of the material, now expressed both in terms of local thresholds (imposed quenched heterogeneity) and elastic properties (emerging and evolving heterogeneity), with both components playing a role on the effective temperature through an evolution of the excitation spectrum. In athermal simulations, the consequence is a stronger localization of damage as approaching the peak stress and a more brittle macroscopic behavior \citep{amitrano1999}. 

\subsubsection{Variability of the creep lifetime for given external conditions}\label{chap_numres02}

We now raise the question of the distribution of lifetimes, and of the variability around the mean, for given external conditions (load and thermodynamic temperature) and level of heterogeneity, but different realizations of material heterogeneity and thermal disorder. 
In Sections \ref{chap_analyt_tf_Th} and \ref{chap_analyt_tf01}, we mentioned that, if one assumes the \textit{independence} of the successive transitions $j+1$ leading to creep failure, the distribution of lifetimes should converge towards a Gaussian distribution~(\ref{chap_analyt_tf_Th}) and the associated standard deviation should scale as $\delta t_f = \langle t_{f} \rangle / \sqrt{\mathcal{N}_f}$ (\ref{chap_analyt_tf01}). To what extent this holds for more realistic cases, taking into account elastic stress redistribution and memory effects ?

\begin{figure}[t!]
\centering
\includegraphics[width=\columnwidth]{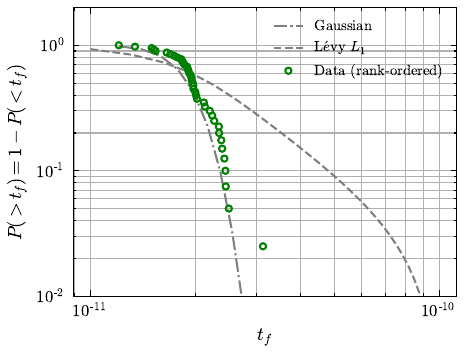}
\caption{Log-log plot of the complementary cumulative distribution of the failure time $P(>t_f)$ for a very heterogeneous material, $\delta C = 5.0$, $T=1000 \; \si{\kelvin}$, $\sigma_{appl}/\sigma_{ath}=0.81$ and 40 realizations of the microstructure (green dots) contrasted to a Gaussian and Lévy distributions.}
\label{fig:num07}
\end{figure}

We used rank-ordered statistics to identify if the mean $\langle t_f \rangle$ and the standard deviation $\delta t_f$ of the creep lifetimes are well-defined and do not follow a (Lévy) power-law distribution whose statistics are more complicated as they depend on the sampling size \citep{sornette2006}. The Fig. \ref{fig:num07} represents
the rank-ordered failure times~\citep{sornette2006} for very heterogeneous samples under the same initial conditions (green dots). 
This is contrasted with a Gaussian complementary cumulative distribution (gray dashed line with points) with the same mean value and standard deviation, and with a complementary cumulative Lévy distribution characterized by a fat tail, $L_1(>t_f)\sim t_f^{-2}$ \citep{sornette2006} (gray dashed line). This shows that the empirical distribution of failure times~$P(<t_f)$ only slightly deviates from the Gaussian distribution, especially in the tail. Such a deviation is not surprising in view of the fact that the successive transitions $j$ are not independent as the result of stress redistribution and memory effects. However, the empirical tail decays much faster than~$t_f^{-2}$, i.e. the lifetimes distribution falls within the Gaussian attraction basin \citep{sornette2006}, and the mean $\langle t_f \rangle$ and the standard deviation $\delta t_f$ are well-defined.  

\begin{figure}[t!]
    \begin{subfigure}[b]{\columnwidth}
        \centering
        \includegraphics[width=\columnwidth]{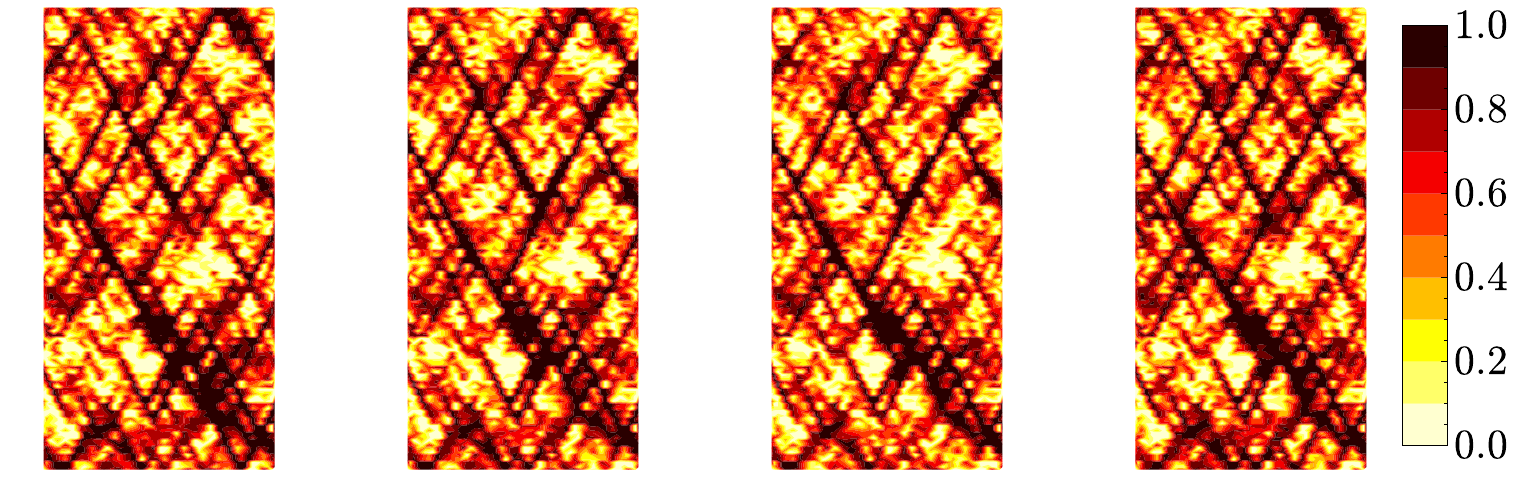}
        \caption{\label{fig:num06a} Very heterogeneous system.}
    \end{subfigure}
    \begin{subfigure}[b]{\columnwidth}
        \centering
        \includegraphics[width=\columnwidth]{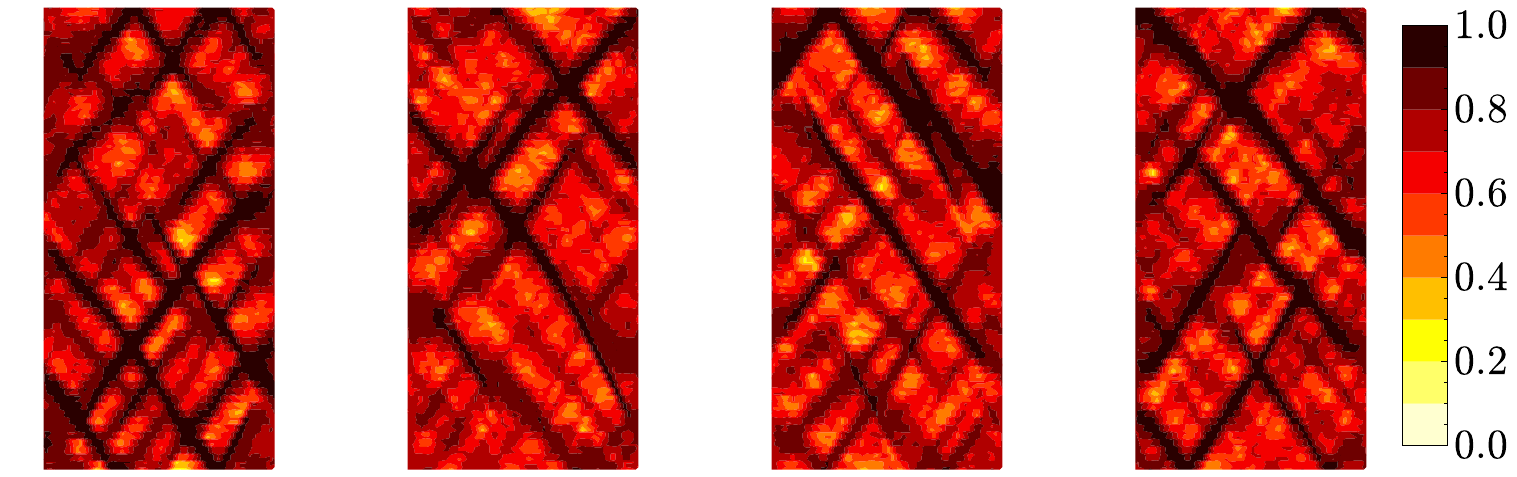}
        \caption{\label{fig:num06b} Homogeneous system.}
    \end{subfigure}
    \caption{\HL{Spatial distribution of the accumulated damage field at failure for a) one very heterogeneous system and b) one homogeneous system, when the KMC algorithm is launched four different times. $T=1000 \; \si{\kelvin}, \; \sigma_{appl}/\sigma_{ath}=0.81 \text{ and } \mu = 0.7$.}}
    \label{fig:num06}
\end{figure}

\begin{figure}[t!]
\centering
\includegraphics[width=\columnwidth]{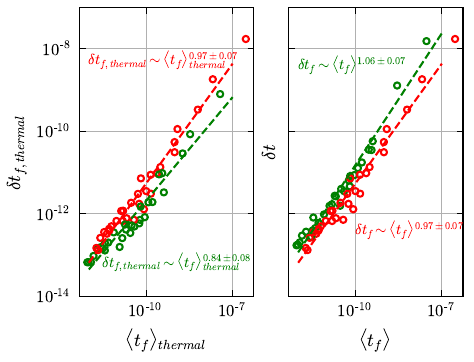}
\caption{Dependence of the standard deviation on the mean failure time. Each dot corresponds to a set of experiments under the same load ratio, temperature and $\mu = 0.7$. \HL{Red: $\delta C = 0$, and green: $\delta C = 5$. Different realizations of the thermal disorder on a system with a same initial quenched heterogeneity (left) and a variable initial quenched heterogeneity (right).}}
\label{fig:num08}
\end{figure}

For a given "material" (given initial distribution of cohesion) and fixed external conditions (temperature and load), the variability of lifetimes can originate from two different forms of stochasticity, the thermal disorder in one hand, and the spatial disorder (initial material heterogeneity) in the other hand, $\delta t_f = f(\delta t_{f,thermal}, \delta t_{f,spatial})$. 
Considering first initially homogeneous systems, we showed that the entropy of the system is maximized, with $S_{homog} = k_B\log{N}$ (see appendix \ref{chap_analyt_thermo}). In this case, the variability of lifetimes obviously comes from thermal disorder only, i.e. $\delta t_f = \delta t_{f,thermal}$ and is associated to a maximum spatial unpredictability (see Fig.~\ref{fig:num06b}).
On the reverse, for simulations performed for a strong heterogeneity, different thermal disorders but with strictly the same initial cohesion field (identical samples), we show in Fig.~\ref{fig:num06a} that the resulting damage patterns at failure are strongly similar. In this case, the system entropy is small, the rupture is no more controlled by thermal activation, but by the initial quenched heterogeneity. In other words, it becomes "deterministic", as long, of course, one has a perfect knowledge of the initial microstructure. Additionally, Fig.~\ref{fig:num08} show that the effect of thermal disorder on the creep lifetime variability, $\delta t_{f,thermal}$, decreases as the heterogeneity increases (see Fig.~\ref{fig:num08} on the left) while this lifetime variability is, overall, larger in heterogeneous samples compared to homogeneous ones, as the result of adding the spatial disorder effect, $\delta t_{f,spatial}$ (see Fig.~\ref{fig:num08} on the right).

Finally, we show in Fig.~\ref{fig:num08} the correlation between the standard deviation and the mean of the creep lifetime.  
In all cases, for both homogeneous and strongly heterogeneous samples, the lifetime variability is essentially proportional to the mean, $\delta t_f \sim \langle t_f \rangle$
, in agreement with our former theoretical expectation. This means that the non-independence of the successive transitions, resulting from elastic interactions and memory effects, has only a limited effect on this scaling, and shows that lifetime variability increases, like the mean lifetime, with decreasing temperature $T$, decreasing applied stress, decreasing quenched heterogeneity, and the nature of the elastic interaction kernel, the two last effects through an effective temperature $T_{eff}$. 

\subsubsection{Implications for the estimation of the activation volume}\label{chap_numres03}

Representing creep-related macroscopic variables, such as the time to failure or the minimum ("secondary") strain-rate, as a function of thermodynamic temperature~$T$ on an Arrhenius plot is a usual practice to estimate an activation energy and/or an activation volume~$V_a$, and consequently to identify an underlying microscopic mechanism controlling creep deformation. This is done while ignoring the effect of microstructural heterogeneity on the relevant effective temperature $T_{eff}$, and thus can lead to false estimations of the activation energy or activation volume.

To illustrate this, we analyze below the implication on the estimation of the activation volume if one assumes that the failure time of a brittle sample submitted to a constant load follows a "classical" Arrhenius law of the form $t_f \sim \exp{(V_{a,est} \Delta\sigma_{macro} / k_BT)}$, i.e.~the activation volume $V_{a,est}$ is estimated from $V_{a,est} \approx \partial\log{t_f} / \partial( \Delta\sigma_{macro}/ k_BT)$, where $\Delta\sigma_{macro} = (\sigma_{ath} - \sigma_{appl})(1-\sin\phi)/2$ is a stress gap at macroscale between a (supposedly known) athermal strength and the applied stress obtained from the Mohr--Coulomb rupture criterion at macroscale (See Fig.~\ref{fig:pdm02}), i.e.~$\sigma_{3, ath} = \sigma_{3, app} = 0$, $\sigma_{1, ath} = \sigma_{ath}$ and $\sigma_{1, app} = \sigma_{app}$. Now, taking into account the heterogeneity effect, we know that the failure time actually follows an Arrhenius-like relation of the form $t_f \sim \exp{(V_{a} \langle \Delta\sigma \rangle_0 / k_BT_{eff})}$, which considers the true activation volume $V_{a}$. 
From this, we can write:
\begin{equation}\label{eqres004}
\frac{V_{a,est}}{V_{a}} \approx \frac{T}{T_{eff}} \frac{\langle \Delta\sigma \rangle_0}{\Delta\sigma_{macro}},
\end{equation}
and replacing the empirical expression for the effective temperature (Eq. \ref{eqres001}):
\begin{equation*}
\frac{V_{a,est}}{V_{a}} \approx \left[ 1 - \alpha (1+\sin\phi) \frac{\delta_{\Delta\sigma,0}}{\langle \Delta \sigma \rangle_0} \right] \frac{\langle \Delta\sigma \rangle_0}{\Delta\sigma_{macro}},
\end{equation*}
or:
\begin{equation}\label{eqres005}
V_{a} \approx V_{a,est} \frac{\Delta\sigma_{macro}}{ \langle \Delta \sigma \rangle_0 - \alpha (1+\sin\phi)\delta_{\Delta\sigma,0}}.
\end{equation}

\begin{figure}[h]
\centering
\includegraphics[width=\columnwidth]{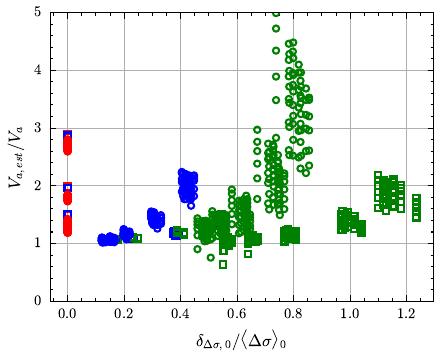}
\caption{\HL{Dependency of the activation volume ratio~$V_{a,est} / V_a$ on the heterogeneity ratio $\delta_{\Delta\sigma,0}/\langle \Delta\sigma \rangle_0$.}}
\label{fig:num05}
\end{figure}

This indicates that the estimated value of the activation volume, $V_{a,est}$, almost systematically overestimates the true microscopic value $V_a$, from a combined effect of quenched and emerging (damage) heterogeneity. Fig.~\ref{fig:num05}, representing the ratio $V_{a,est} / V_a$ for all our creep simulations as a function of the quenced heterogeneity $\delta_{\Delta\sigma,0}/\langle \Delta\sigma \rangle_0$, confirms this point. In particular, $V_{a,est}/V_a>1$, even for initially homogeneous systems, as the result of an emerging heterogeneity of the damage field.

\section{Summary and conclusions}

Previous studies, based on simple fiber-bundle models of brittle creep with a mean-field elastic stress redistribution kernel, suggested that the material microstructural heterogeneity amplifies, for a given external load, the thermodynamic temperature $T$ , i.e. shortens creep lifetimes, an effect that can be interpreted in an Arrhenius formalism from an effective-temperature $T_{eff}$ \citep{ciliberto2001,roux2000}. Here we extended these former works to more complex (and realistic) situations taking into account non-mean-field and non-convex elastic redistribution kernels as well as cumulative memory effects, on the basis of 
a numerical progressive damage model incorporating thermal activation. 
We first showed, from a theoretical analysis and for an initially heterogeneous material while neglecting elastic stress redistribution and memory effects in the course of creep deformation, that the average creep lifetime still follows an Arrhenius-like expression, however with an effective temperature explicitly increasing with the heterogeneity of the material as found by Ref. \citep{ciliberto2001,roux2000}. As shown by our numerical study, this holds true qualitatively as well in the most general case. 
However, in that case, the effective temperature depends both on the level of heterogeneity and on the nature of the elastic interaction kernel, with an increasing convexity enhancing the effective temperature.

Our analysis also showed that the variability $\delta t_f$ of lifetimes around its mean, for fixed thermodynamic temperature, load and level of heterogeneity, is essentially proportional to this average lifetime $\langle t_f \rangle$.
This means that the non-independence of the successive transitions, resulting from elastic interactions and memory effects, has only a limited effect on this scaling. This implies that this variability follows the same dependence with the effective temperature, and therefore with the level of heterogeneity or the nature of the elastic kernel. It means also that, for highly heterogeneous materials, $\delta t_f$ mainly arises from sample-to-sample fluctuations of the initial heterogeneity, and not on thermodynamic disorder.

Finally, we discussed the implication of these results in the estimation of microscopic parameters, such as an activation volume, from macroscopic data obtained from creep tests performed at different \textit{thermodynamic} temperatures $T$. In particular, we showed that using such downscaling procedure while ignoring these effective temperature effects can lead to a strong overestimation of the real activation volume at the microscale.

It would be interesting to look the possible implications on the creep lifetime distribution for different kind of initial heterogeneity distribution, e.g. Weibull distribution. Additionally, concerning the activation volume, the results show that a great caution should be taken when trying to estimate a microscopic activation energy or volume by plotting creep macroscopic variables within a classical Arrhenius plot while ignoring the potential effect of microstructural heterogeneity. 

\begin{acknowledgments}
ISTerre is part of Labex OSUG@2020. This work has been supported by the French National Research Agency in the framework of the "Investissements d'Avenir" program (ANR-15-IDEX-02). We thank Alberto Rosso for interesting discussions. ISTerre IT is acknowledged for computational resources. 
J.C.V.E, T.M. and M.A. acknowledge the support from FinnCERES flagship (grant no. 151830423), Business Finland (grant nos. 211835, 211909, and 211989), and Future Makers programs. 
M.A. acknowledges support from the Academy of Finland (\HL{grant nos. 361245} and 317464), as well as the Finnish Cultural Foundation. 
M.A. acknowledges support from the European Union Horizon 2020 research and innovation program under Grant Agreement No. 857470 and the European Regional Development Fund via the Foundation for Polish Science International Research Agenda PLUS program under Grant No. MAB PLUS/2018/8.
\end{acknowledgments}


\bibliography{apssamp}

\newpage
\newpage

\appendix

\section{Theoretical analysis}\label{app:a0}

\subsection{Spatial heterogeneity analysis}\label{chap_analyt_thermo}

Spatial heterogeneity of the material is introduced at the scale of the subvolumes, each one associated to a variable activation energy $E_{n,j}$ or energetic state, which is also evolving with the transition number $j$ as the result of previous transitions and their associated stress redistributions (see Fig.~\ref{fig:entropy00}).

As we consider a mechanical system, we link the activation energy $E_{n,j}$ to a local stress gap $\Delta\sigma_{n,j}$, i.e. $E_{n,j}= \Delta\sigma_{n,j} V_a$, where $V_a$ is the activation volume for the considered damage/fracturing microscopic mechanism, considered here as being constant in space and time. The stress gap $\Delta\sigma_{n,j}$ represents a distance, expressed in the principal stress space, between the local  threshold $\sigma_{ath,n}$ (hereafter denoted athermal as it corresponds to a failure threshold in absence of thermal disorder) and the local stress state, $\Delta\sigma_{n,j} \sim \sigma_{ath,n} - \sigma_{n,j}$, here we do not address how the stress on each element may vary in space, democratic hypothesis leads to consider that stress is equal in each element. Material heterogeneity is introduced at the level of each subvolume $n$ through $\sigma_{ath,n}$ considered as a quenched random variable~\citep{makinen2023history, weiss2023logarithmic}.

The goal is to understand how the coupling between the material heterogeneity and the (thermodynamic) disorder affects the creep rupture time $t_f$ in heterogeneous systems. We first analyze the general case without any assumptions concerning the analytical form of the material heterogeneity,

Recalling the definition of the average waiting time between transitions, $\langle \Delta t \rangle_{j+1} \sim \exp{(F_{j}/k_BT)}$, we realize that an important variable of our problem is the free energy $F_j$ of the system. $F_j$ is defined as the difference between the average energy weighted by the activation energy probability $p(E_{n_b,j})$ (at the transition $j+1$), $\langle E \rangle_j$, and the term linked to the entropy~$TS_j$~\citep{sornette2006, pathria2011, sethna2021entropy}:
\begin{eqnarray}\label{eqtheorApp001e}
F_j = \langle E \rangle_j - TS_j = -k_BT\log{Z_j}.
\end{eqnarray}

It is important to mention that the thermodynamical variables showed in the following are not interpreted in the sense of the kinetic theory of gases, but in that of the Shannon's theory of (lack of) information \citep{shannon1948, sornette2006, pathria2011, sethna2021entropy}. In this context, the Shannon entropy times the Boltzmann constant $k_B$ is given by:
\begin{eqnarray}\label{eqshannonApp}
S_j= - k_B\sum_n p(E_{n,j})\log{p(E_{n,j})},
\end{eqnarray}
where the probability of a solid composed of $N$ sub-volumes to be in a configuration with energy $E_{n,j}$ is $p(E_{n,j}) = e^{-E_{n,j}/k_B T}/Z_j$~\citep{fichthorn1991, sornette2006, pathria2011, sethna2021entropy}, where $Z_j$ is the partition function \citep{sornette2006, pathria2011, sethna2021entropy} defined by:
\begin{eqnarray}\label{eqtheorApp001b}
Z_j = \sum_n e^{-E_{n,j}/k_B T}.
\end{eqnarray}

As an alternative representation of the last equations, it is more convenient to introduce the density of states $g(E_{n_b,j})$, which corresponds to the number of energetic states, i.e. the number of the sub-volumes, that have an energy within a small range $E_{n_b,j} \pm dE_{n_b,j}$, with $n_b$ representing the number of the bin in which the energy value $E_{n_b,j}$ falls  \citep{sornette2006, pathria2011}. We can then rewrite for a given bin $n_b$, $g(E_{n_b,j}) dE_{n_b,j} = N f(E_{n_b,j}) dE_{n_b,j}$, which can be replaced into the partition function \citep{pathria2011}:
\begin{eqnarray}\label{eqtheorApp001c}
Z_j = N \sum_{n_b} f(E_{n_b,j}) e^{-E_{n_b,j}/k_B T},
\end{eqnarray}
 where $f(E_{n_b,j}) = g(E_{n_b,j}) / N \approx f(\Delta\sigma_{n_b,j}) / V_a$ represents a relative density of states and is an explicit representation of the spatial heterogeneity in the system. It is equivalent to the probability distribution of the stress gap (also called the excitation spectrum), taking into account that the activation volume is assumed constant here. The probability that the system is in a specific state with energy 
$E_{n_b,j}$ is:
\begin{eqnarray}\label{eqtheor001d}
p(E_{n_b,j}) = \frac{Nf(E_{n_b,j})e^{-E_{n_b,j}/k_B T}}{Z_j}.
\end{eqnarray}

The average energy of the system $\langle E \rangle_j$ at transition $j+1$ is the arithmetic mean value of the energy barrier weighted by the activation energy probability $p(E_{n_b,j})$. It differs from $\langle E_{\Delta\sigma}\rangle_j$, which corresponds to the "classical" arithmetic mean value of the energy. Then, the average energy of the system is given by:
\begin{eqnarray*}\label{eqtheor001f}
\langle E \rangle_j = \sum_{n} E_{n,j} p(E_{n,j}) = \frac{N}{Z_j}\sum_{n_b} E_{n_b,j} f(E_{n_b,j}) e^{-E_{n_b,j}/k_B T},
\end{eqnarray*}
and, using the expression of $Z_j$ in Eq.~\ref{eqtheorApp001c}:
\begin{eqnarray}\label{eqtheor001g}
\langle E \rangle_j = \frac{\sum_{n_b} E_{n_b,j} f(E_{n_b,j}) e^{-E_{n_b,j}/k_B T}}{\sum_{n_b} f(E_{n_b,j}) e^{-E_{n_b,j}/k_B T}} .
\end{eqnarray}
Solving for the entropy term in Eq.~\ref{eqtheorApp001e}, $TS_j = \langle E \rangle_j + k_BT\log{Z_j}$, one obtains:
\begin{multline}\label{eqtheor001h}
S_j = k_B \log{N} + \frac{1}{T}\frac{\sum_{n_b} E_{n_b,j} f(E_{n_b,j}) e^{-E_{n_b,j}/k_B T}}{\sum_{n_b} f(E_{n_b,j}) e^{-E_{n_b,j}/k_B T}}   \\
+ k_B \log{\left[\sum_{n_b} f(E_{n_b,j}) e^{-E_{n_b,j}/k_B T}\right]}, 
\end{multline}
which could be rewritten as:
\begin{multline}\label{eqtheor001i}
S_j = k_B \log{ \left[ N \sum_{n_b} f(E_{n_b,j}) \exp{\left( -\frac{E_{n_b,j} - \langle E \rangle_j}{k_B T} \right)} \right]}  \\
= k_B \log{ \left[ N \Biggl \langle \exp{\left( -\frac{E_{n_b,j} - \langle E \rangle_j}{k_B T} \right)} \biggr \rangle \right]}.
\end{multline}

Equation \ref{eqtheor001i} represents the entropy as a function of any density of states. Note that this expression could have been obtained from Eq.~\ref{eqshannonApp}, using a more involved derivation. A first key result is that the material heterogeneity of a system can indeed affect its entropy. As the entropy is $k_B\text{log}(\Omega)$ where $\Omega$ is the number of possible microstates, this number reads, from Eq.~\ref{eqtheor001i}, $\Omega=N \langle \exp{[ - (E_{n_b,j} - \langle E \rangle_j) / k_B T ]} \rangle$ \citep{pathria2011, sethna2021entropy}. One obvious remark is that, for the case of a perfectly homogeneous system, i.e.   $p(E_{n,j}) = 1 / N$ and $E_{n,j} = \langle E \rangle_j$, or when the thermodynamic temperature is very large, $T \to \infty$, the entropy reaches a maximum value $S_j = k_B \log{N}$ which describes a system with all the energetic states having the same probability of occurrence. This represents the limit of infinite (thermal) disorder \citep{sornette2006}, or a high degree of unpredictability \citep{pathria2011}, in the sense that thermal noise dominates the behavior.

\subsection{Prediction of the rupture time $t_{f}$ and the effective temperature $T_{eff}$ for a heterogeneous system}\label{chap_analytApp_tf}

We switch to the main question of this work: What is the combined effect of material heterogeneity (variability of the stress gaps $\Delta \sigma_j)$ and (thermal) disorder  on creep lifetimes? And, do we still get an Arrhenius-like expression for $t_f$, possibly with an heterogeneity-dependent effective temperature $T_{eff}$?

From the definition of the rupture time  (Eq.~\ref{eqtheor000e}), we can rewrite the expression as:
\begin{eqnarray}\label{eqtfApp001}
\langle t_f \rangle = \frac{1}{\Omega_0} \sum_{j=0}^{\mathcal{N}_{f}-1} \exp{\left(\frac{F_{j}}{k_B T} \right)} = \mathcal{N}_{f} \overline{\langle \Delta t \rangle}_{j+1}
\end{eqnarray}
where $\overline{\langle \Delta t \rangle}_{j+1}$ the arithmetic mean value of all the time intervals between successive transitions along the experiment and $\mathcal{N}_{f}$ is the total number of these transitions until failure, considered as being constant in what follows. As a remark, for a case for which each element $n$ can be only activated once, such as the FBM studied by Ciliberto et al. \cite{ciliberto2001} and Roux \cite{roux2000}, this maximum number of transitions equals the number of \HL{sub-volumes} $\mathcal{N}_{f} = N$. 

The combination of Eqs. \ref{eqtheorApp001e}, \ref{eqtheor001g}, \ref{eqtheor001h} and \ref{eqtfApp001} shows that (i)~the free energy $F_{j}$ evolves during creep and (ii)~the failure time depends on the initial quenched heterogeneity, on the applied stress through the distribution of stress gaps $\Delta\sigma_{j}$, as well as on thermal fluctuations $k_BT$. Moreover, the mechanical interactions between neighboring sub-volumes after each transition also modify the stress gaps distributions.

In order to obtain an expression for the effective temperature $T_{eff}$, we define first the term~$\Delta t_{h,1}$ corresponding to an initial time step for an "equivalent" homogeneous system, i.e.~with maximum entropy, $S_{h,0} = k_B\log{N}$ and the average energy equalling the arithmetic mean energy barrier $\langle E\rangle_{h,0}=\langle E_{\Delta\sigma}\rangle_{0}=E_{\Delta \sigma_{h,0}}=V_a\Delta\sigma_{h,0}$. Consequently, 
\begin{eqnarray}\label{eqtfApp003a}
\Delta t_{h,1} = \Omega_0^{-1} \exp{\left(\frac{F_{h,0}}{k_BT}\right)}=\frac{1}{N\Omega_0}\exp{\left(\frac{E_{\Delta \sigma_{h,0}}}{k_B T} \right)},
\end{eqnarray}
which follows an Arrhenius law and decreases with the system size $N$. Combining this with Eq.~\ref{eqtfApp001}, we obtain:
\begin{eqnarray*}
\langle t_f \rangle = \frac{1}{\Omega_0}\frac{\mathcal{N}_{f}}{N}\exp{\left(\frac{ E_{\Delta \sigma_{h,0}}}{k_B T} \right)} \frac{\overline{\langle \Delta t \rangle}_{j+1}}{\Delta t_{h,1}},
\end{eqnarray*}
and reorganizing:
\begin{eqnarray*}
\langle t_f \rangle = \frac{\mathcal{N}_{f}}{N\Omega_0}\exp{\left[ \frac{ E_{\Delta\sigma_{h,0}}}{k_B T}  \left( 1 + \frac{k_BT}{E_{\Delta \sigma_{h,0}}} \log{\left(\frac{\overline{\langle \Delta t \rangle}_{j+1}}{\Delta t_{h,1}} \right)} \right) \right]},
\end{eqnarray*}
we can identify an Arrhenius-like expression for the average failure time $\langle t_f \rangle = t_0 \exp{(E_{\Delta \sigma_{h,0}}/k_BT_{eff})}$, with the effective temperature $T_{eff}$ and the "characteristic" timescale $t_0$ of equation \ref{eqintro001} given respectively by:
\begin{eqnarray}\label{eqtfApp003}
k_B T_{eff} = \frac{k_B T}{ 1 - \Delta F_{cor} / E_{\Delta \sigma_{h,0}}}, \\
t_0 = \frac{1}{\Omega_0}\frac{\mathcal{N}_{f}}{N},
\end{eqnarray}
where the term $\Delta F_{cor}$ corresponds to a corrected differential free energy given by: 
\begin{multline}\label{eqtf004}
\Delta F_{cor} = k_B T \log{\left( \frac{\Delta t_{h,1}}{\overline{\langle \Delta t \rangle}_{j+1}} \right)} \\
= - k_BT \log{ \left[ \frac{1}{\mathcal{N}_{f}} \sum_{j=0}^{\mathcal{N}_{f}-1} \exp{\left(-\frac{\Delta F_{j}}{k_B T} \right)} \right]} \\
= - k_BT \log{ \left[  \overline{\exp{\left(-\frac{\Delta F_{j}}{k_B T} \right)}} \right]},
\end{multline}
and $\Delta F_{j} = F_{h,0} - F_{j} \geq 0$ is the differential free energy between the "equivalent" homogeneous free energy at the beginning of the experiment, $F_{h,0} = E_{\Delta \sigma_{h,0}} - k_BT\log{N}$, and the free energy after a transition $j$ is triggered by thermal activation, $F_{j}$. $\Delta F_{j}$ is a measure of the heterogeneity of a system, just before its transition $j$.

\subsection{Lifetime for a Gaussian distribution of the stress gap}\label{app:a}

We consider that the activation energy, $E_{j} = V_a\Delta\sigma_{j}$, at a given $j$th transition follows a Gaussian density of states $E_{j}\sim\mathcal{N}(\langle E_{\Delta\sigma}\rangle_j,\delta E_{\Delta\sigma,j}^2)$ with arithmetic mean value and variance $\langle E_{\Delta\sigma}\rangle_j$, $\delta E_{\Delta\sigma,j}^2$ respectively, then the probability function $p(E_{\Delta\sigma, j})$ is defined by:
\begin{eqnarray}\label{eqappA01}
f(E_{\Delta\sigma, j}) = \frac{1}{\delta E_{\Delta\sigma,j}\sqrt{2\pi}} \exp{\left[ -\frac{1}{2} \left( \frac{E_j - \langle E_{\Delta\sigma}\rangle_j}{\delta E_{\Delta\sigma,j}}  \right)^2 \right]}.
\end{eqnarray}

Additionally, the average weighted energy $\langle E \rangle_j$ and the entropy $S_j$ if the number of sub-volumes tends to infinity, $N \to \infty$, are given by:
\begin{equation}\label{eqappA02}
\langle E \rangle_j = \frac{\int_0^{\infty} E_j f(E_{\Delta \sigma,j}) e^{-E_j/k_B T} dE_j}{\int_0^{\infty} f(E_{\Delta \sigma,j}) e^{-E_j/k_B T} dE_j},
\end{equation}
\begin{multline}\label{eqappA03}
S_j = k_B \log{N} + \frac{1}{T}\frac{\int_0^{\infty} E_j f(E_{\Delta \sigma,j}) e^{-E_j/k_B T} dE_j}{\int_0^{\infty} f(E_{\Delta \sigma,j}) e^{-E_j/k_B T} dE_j}   \\
+ k_B \log{\left[\int_0^{\infty} f(E_{\Delta \sigma,j}) e^{-E_j/k_B T} dE_j\right]}.
\end{multline}

Using some variable changes and solving the integrals by parts we obtain:
\begin{multline}\label{eqappA06}
\langle E \rangle_j = \langle E_{\Delta\sigma}\rangle_j - \frac{\delta E_{ \Delta\sigma,j}^{2}}{k_BT} + \sqrt{\frac{2}{\pi}} \delta E_{ \Delta\sigma,j} \frac{e^{-(z_{var,j}/\sqrt{2})^2}}{\erfc{[z_{var,j}/\sqrt{2}]}},
\end{multline}
\begin{multline}\label{eqappA07}
S_j = k_B \log{\left[ \frac{N}{2} \erfc{\left( \frac{z_{var,j}}{\sqrt{2}} \right)} \right]} - k_B \left(\frac{\delta E_{\Delta\sigma,j}}{k_BT} \right)^2\\
+ \sqrt{\frac{2}{\pi}} \frac{\delta E_{ \Delta\sigma,j}}{T} \frac{e^{-(z_{var,j}/\sqrt{2})^2}}{\erfc{[z_{var,j}/\sqrt{2}]}},
\end{multline}
with $z_{var,j} = \delta E_{\Delta\sigma,j}/k_BT -\langle E_{\Delta\sigma}\rangle_j / \delta E_{\Delta\sigma,j}$ and $\erfc{(\textbf{x}^*)}=2/\sqrt{\pi}\int_{\textbf{x}^*}^{\infty} e^{-x^2}  dx$, the Gauss complementary error function. Consequently the free energy, $F_j = \langle E \rangle_j - TS_j$ is
\begin{multline}\label{eqappA08}
F_j = \langle E_{\Delta\sigma}\rangle_j - k_BT \log{\left[ \frac{N}{2} \erfc{\left( \frac{z_{var,j}}{\sqrt{2}} \right)} \right]} - \frac{1}{2} \frac{\delta E_{\Delta\sigma,j}^2}{k_BT}.
\end{multline}

Taking into account that the lifetime of a single $j$th transition is given by:
\begin{equation}\label{eqappA09}
\langle \Delta t \rangle_j = \frac{1}{\Omega_0} e^{F_j / k_B T},
\end{equation}
we can obtain an expression for the lifetime of a single transition by substituting Eq.~\ref{eqappA08} into Eq.~\ref{eqappA09},
\begin{multline}\label{eqappA10}
\langle \Delta t \rangle_j = \frac{2}{N\Omega_0} e^{\langle E_{\Delta\sigma,j} \rangle_j / k_B T} e^{-\frac{1}{2} \left(\delta E_{\Delta\sigma,j} / k_BT \right)^2} \times \\
\left[\erfc{\left( \frac{z_{var,j}}{\sqrt{2}} \right)} \right]^{-1},
\end{multline}
which in the case of negligible mechanical interactions is constant for all transitions. This corresponds to a Poisson process, and consequently the failure time after $\mathcal{N}_{f}$ transitions is:
\begin{multline}\label{eqappA11}
\langle t_{f} \rangle^{gauss} \approx \frac{2}{\Omega_0} \frac{\mathcal{N}_{f}}{N} e^{\langle E_{\Delta\sigma,j} \rangle_0 / k_B T} e^{-\frac{1}{2} \left(\delta E_{\Delta\sigma,0} / k_BT \right)^2} \times \\
\left[\erfc{\left( \frac{z_{var,0}}{\sqrt{2}} \right)} \right]^{-1},
\end{multline}
which could be re-written as $\langle t_{f} \rangle^{gauss} = t_0 e^{\langle E_{\Delta\sigma,j} \rangle_0 / k_B T_{eff}}$, with $t_0 = \mathcal{N}_{f} / N\Omega_0$ and $T_{eff}$ is an effective temperature given by:
\begin{equation}\label{eqappA12}
k_BT_{eff} \approx \frac{k_BT}{1 - \Delta F / \langle E_{\Delta\sigma} \rangle_0},
\end{equation}
with $\Delta F$ a differential free energy between the free energy at the first transition, $F_0$, and the free energy of a homogeneous equivalent system, $F_{h,0}=\langle E_{\Delta\sigma,j} \rangle_0 - k_BT\log{N}$. Then the effective temperature could be re-written as:
\begin{equation}\label{eqappA12}
k_BT_{eff} \approx \frac{k_BT}{1 - \frac{k_BT}{\langle E_{\Delta\sigma} \rangle_0}\log{\left[ \frac{1}{2} e^{\frac{1}{2} \left(\delta E_{\Delta\sigma,j} / k_BT \right)^2} \erfc{\left( \frac{z_{var,j}}{\sqrt{2}} \right)} \right]}}.
\end{equation}

\subsection{Lifetime for a uniform distribution of the stress gap}\label{app:b}

We now consider that at a given $j$th transition, the activation energy follows a uniform density of states, $E_j \sim \mathcal{U}_{[E_{j,min}, E_{j,max}]}$ with maximum and minimum values $V_a\Delta\sigma_{max}$ and $V_a\Delta\sigma_{min}$. In this case, the arithmetic mean and corresponding variance are $\langle E_{\Delta\sigma}\rangle_j$, $\delta E_{\Delta\sigma,j}^2$ respectively, then the probability function $f(E_{\Delta\sigma,j})$ is defined by: 
\begin{eqnarray}\label{eqappB01}
f(E_{\Delta\sigma,j}) = \frac{1}{E_{j,max}-E_{j,min}} = \frac{1}{2\Delta E_j},
\end{eqnarray}
with $\Delta E_j = \sqrt{3}\delta E_{\Delta\sigma,j}$. Using some variables change and solving the integrals by parts we obtain:
\begin{equation}\label{eqappB04}
\langle E \rangle_j = \langle E_{\Delta\sigma}\rangle_j + k_BT - \Delta E_j \coth{\left( \frac{\Delta E_j}{k_B T} \right)} ,
\end{equation}
\begin{multline}\label{eqappB05}
S_j = k_B \log{\left[ N e \frac{k_BT}{\Delta E_j} \sinh{\left( \frac{\Delta E_j}{k_B T}\right)} \right]} - \frac{\Delta E_j}{T} \coth{\left( \frac{\Delta E_j}{k_B T}\right)},
\end{multline}
and consequently the free energy is
\begin{multline}\label{eqappB06}
F_j = \langle E_{\Delta\sigma}\rangle_j - k_BT \log{\left[ N \frac{k_BT}{\Delta E_j} \sinh{\left( \frac{\Delta E_j}{k_B T}\right)} \right]}.
\end{multline}

\begin{figure}[t!]
    \begin{subfigure}[b]{0.54\columnwidth}
        \centering
        \includegraphics[width=\columnwidth]{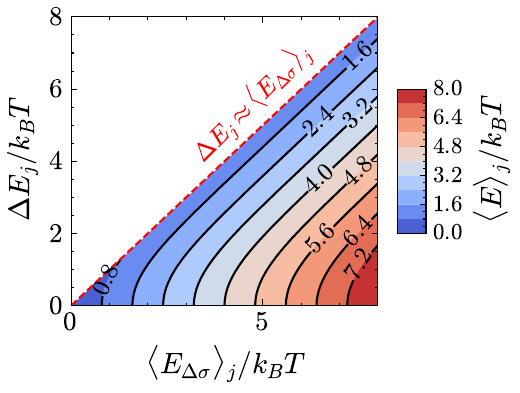}
        \caption{\label{fig:Unif01a} Average energy.}
    \end{subfigure}%
    \begin{subfigure}[b]{0.54\columnwidth}
        \centering
        \includegraphics[width=\columnwidth]{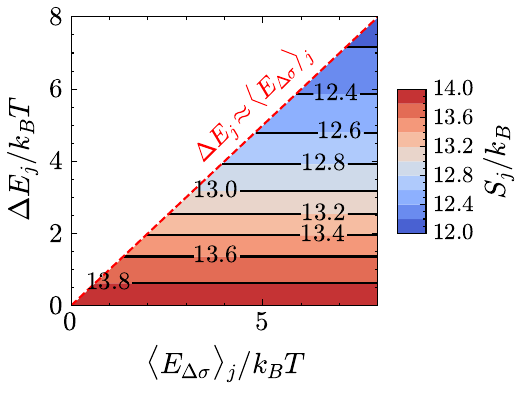}
        \caption{\label{fig:Unif01b} Entropy.}
    \end{subfigure}
    \begin{subfigure}[b]{0.5\columnwidth}
        \centering
        \includegraphics[width=\columnwidth]{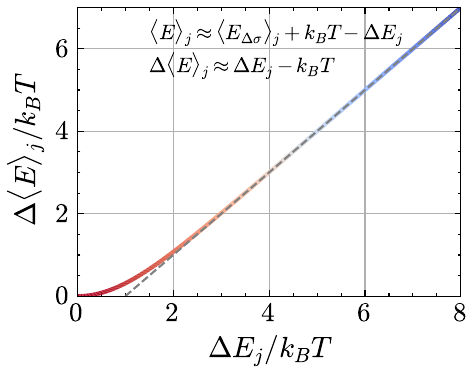}
        \caption{\label{fig:Unif01c} Differential energy.}
    \end{subfigure}%
    \begin{subfigure}[b]{0.5\columnwidth}
        \centering
        \includegraphics[width=\columnwidth]{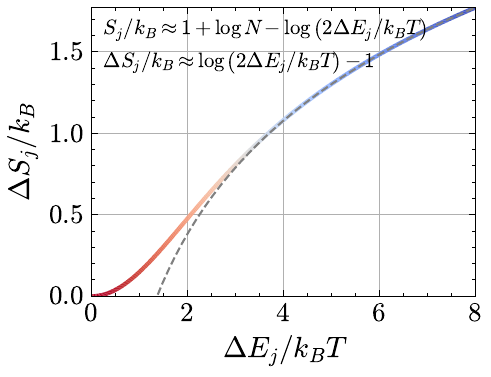}
        \caption{\label{fig:Unif01d} Differential entropy.}
    \end{subfigure}
    \caption{Example of average energy and entropy, both normalized by $k_BT$, as a function of the arithmetic mean value and the difference between the maximum and the minimum value of the activation energy of the initial distribution of the stress gap.}
    \label{fig:Unif01}
\end{figure}
In Figs.~\ref{fig:Unif01a} and \ref{fig:Unif01b} we show respectively the average weighted energy $\langle E \rangle_j$ and the entropy term $TS_j$, as a function of the arithmetic mean value of the local activation energies and a term proportional to the standard deviation $\Delta E_j \sim \delta E_{\Delta\sigma,j}$, both normalized by $k_BT$. From Fig.~\ref{fig:Unif01a} we see that the average weighted energy $\langle E \rangle_{j}$ increases when increasing the arithmetic mean value of the local activation energies $\langle E_{\Delta\sigma} \rangle_j$, with $\langle E \rangle_j \to \langle E_{\Delta\sigma} \rangle_j$ when the system tends to be homogeneous, i.e.~$\Delta E_j \to 0$. On the reverse, for very large standard deviations, i.e.~$\delta E_{\Delta\sigma,j} =\Delta E_j / \sqrt{3} \gg 0$, the average energy vanishes. One additional remark is when we substract the arithmetic mean value of the activation energy from the average weighted energy (Fig.~\ref{fig:Unif01c}), we collapse the data in one single expression which is a function only of the standard deviation of the activation energy $\Delta \langle E \rangle_j = \langle E \rangle_j - \langle E_{\Delta\sigma} \rangle_j = k_BT - \Delta E_j \coth{ (\Delta E_j/k_B T)}$. From this, we can see that for large heterogeneities, $\Delta E_j \geq 2$, the average weighted energy follows $\langle E \rangle_j \approx \langle E_{\Delta\sigma} \rangle_j + k_BT - \Delta E_j$.

Figure \ref{fig:Unif01b} illustrates the entropy term $TS_j$ normalised by $k_BT$. We see that the entropy of a heterogeneous system that follows a uniform distribution only depends on the heterogeneity term $\Delta E_j$ and not on the arithmetic mean value, which is represented in Fig.~\ref{fig:Unif01d}. We see that for a quasi-homogeneous system, $\Delta E_j \to 0$, the entropy is the largest and approximates $S_j \to k_B\log{N}$, as expected. In addition, as for the Gaussian case, an increase of the heterogeneity reduces the entropy. In Fig.~\ref{fig:Unif01d} we collapse the data in one single expression which is function only of the standard deviation of the activation energy. From this plot we can see that for large heterogeneities, $\Delta E_j \geq 2$, the entropy scales as $S_j \sim - k_B\log{(2 \Delta E_j / k_BT)}$.

Then, the duration of a single transition is given by:
\begin{equation}\label{eqappB07}
\langle \Delta t \rangle_j = \frac{1}{N\Omega_0} \frac{\Delta E_j/k_BT}{\sinh{(\Delta E_j/k_BT)}} e^{\langle E_{\Delta\sigma} \rangle_j / k_B T},
\end{equation}
and the creep lifetime of the full volume, neglecting mechanical interaction, after $\mathcal{N}_{f}$ transitions, is given by:
\begin{equation}\label{eqappB08}
\langle t_{f}\rangle^{unif} = \frac{\mathcal{N}_{f}}{N\Omega_0} \frac{\Delta E_0/k_BT}{\sinh{(\Delta E_0/k_BT)}} e^{\langle E_{\Delta\sigma} \rangle_0 / k_B T}.
\end{equation}

\begin{figure}[t!]
\centering
\includegraphics[width=\columnwidth]{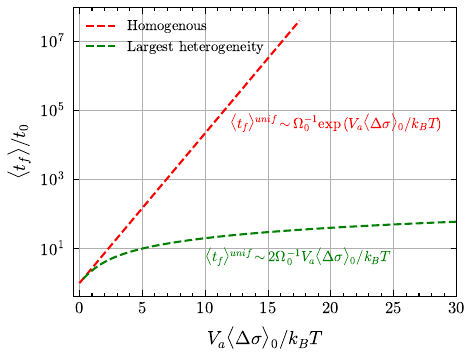}
\caption{\label{fig:Unif02} Dependency of the normalized average lifetime $\langle t_f \rangle / t_0$ on $V_a\langle\Delta\sigma\rangle_0/k_BT$, for an uniform initial distribution of stress gaps, and mechanical interactions neglected. Theoretical result.}
\end{figure}

This shows once again that heterogeneity plays an important role on lifetime. In addition, it is possible to deduce that for large values of temperature or low values of the stress gap, $\langle t_f \rangle^{unif} \sim \Omega_0^{-1}$. For the case of a homogeneous volume ($\Delta E_0 \to 0$), the failure time follows an Arrhenius expression, $\langle t_f \rangle^{unif} \sim \Omega_0^{-1} \exp{\left(\langle E_{\Delta\sigma} \rangle_0 / k_B T \right)}$ as found by Ref.~\citep{zhurkov1965} in his experiments on homogeneous materials. On the other hand, for very large heterogeneities, i.e.~$\Delta E_0 \gg 0$, it is possible to say that $\Delta E_0 \approx \langle E_{\Delta\sigma} \rangle_0$. It implies that Eq.~\ref{eqappB08} can be written as $\langle t_f \rangle^{unif} \sim 2 \Omega_0^{-1} (\langle E_{\Delta\sigma} \rangle_0 / k_BT)  e^{\langle E_{\Delta\sigma} \rangle_0 / k_B T} / (e^{\langle E_{\Delta\sigma} \rangle_0 / k_B T} - e^{-\langle E_{\Delta\sigma} \rangle_0 / k_B T}) \sim 2 \Omega_0^{-1} \langle E_{\Delta\sigma} \rangle_0 / k_BT$ (Fig.~\ref{fig:Unif02}).

In a similar manner as for the Gaussian case, we can re-write the failure time as a function of an effective temperature $T_{eff}$ as $\langle t_f \rangle^{unif} = t_0 e^{\langle E_{\Delta\sigma,j} \rangle_0 / k_B T_{eff}}$, where the effective temperature is given by:
\begin{equation}\label{eqappB09}
k_BT_{eff} \approx \frac{k_BT}{1 - \frac{k_BT}{\langle E_{\Delta\sigma} \rangle_0}\log{\left[ \frac{\sinh{(\Delta E_0/k_BT)}}{\Delta E_0/k_BT} \right]}} .
\end{equation}

\begin{figure}[t!]
\centering
\includegraphics[width=\columnwidth]{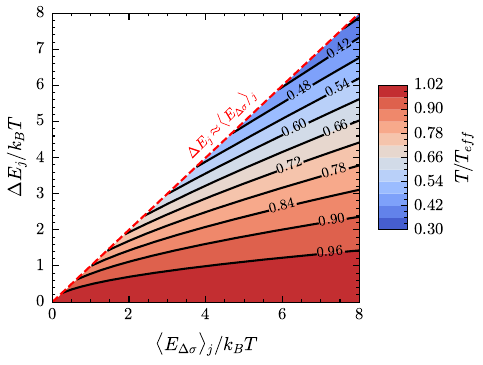}
\caption{\label{fig:Unif03} Example of the effective temperature term $T/T_{eff}$ as a function of the arithmetic mean value and the standard deviation of the initial distribution of the stress gap.}
\end{figure}  

Figure \ref{fig:Unif03} illustrates the temperature ratio $T/T_{eff}$ as a function of the arithmetic mean value of the energy $\langle E_{\Delta\sigma} \rangle_0/k_BT$ and the heterogeneity term $\Delta E_{0}/k_B T$. We can see that for homogeneous systems, $T/T_{eff} \to 1$, as expected. On the contrary, we see that as the heterogeneity increases,  the ratio $T/T_{eff} \to 0$. When $\Delta E_0 \gg 0$, we can rewrite the temperature ratio as:
\begin{equation*}
\frac{T}{T_{eff}} \approx 1 - \frac{\Delta E_0}{\langle E_{\Delta\sigma} \rangle_0}  + \frac{k_BT}{\langle E_{\Delta\sigma} \rangle_0}\log{\left(2 \frac{\Delta E_0}{k_BT} \right)},
\end{equation*}
and for the case of maximum heterogeneity, $\Delta E_0 \approx \langle E_{\Delta\sigma} \rangle_0$, the temperature ratio could be written as:
\begin{equation*}
\frac{T}{T_{eff}} \approx \frac{k_BT}{\langle E_{\Delta\sigma} \rangle_0}\log{\left(2 \frac{\langle E_{\Delta\sigma} \rangle_0}{k_BT} \right)},
\end{equation*}
which converges for largest values of $\langle E_{\Delta\sigma} \rangle_0 / k_BT \to \infty$, $T/T_{eff} \to 0$. 
The results shown in Figs.~\ref{fig:Unif01}, \ref{fig:Unif02} and \ref{fig:Unif03} are very similar to those obtained for a Gaussian distribution of the stress gaps.

\section{Progressive Damage model with thermal activation}\label{app:c}

\subsection{Athermal model}

\begin{figure}[t!]
\centering
\includegraphics[width=\columnwidth]{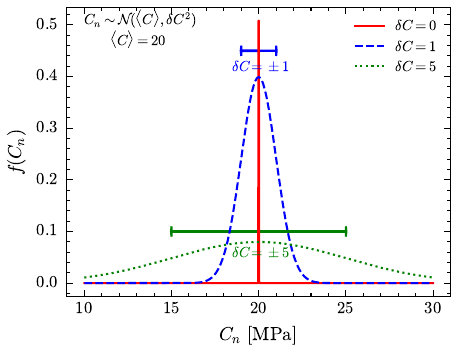}
\caption{\label{fig:pdm01} Gaussian probability distribution of the cohesion for the studied cases. $\langle C \rangle = 20$~\si{\mega\pascal} and $\delta C = 0, 1, 5$~\si{\mega\pascal}.} 
\end{figure}

The athermal version of the progressive damage model has been extensively studied by Refs.~\citep{amitrano1999, amitrano2006, girard2010, girard2012}, and only its main characteristics are recalled below. We consider a 2D elastic domain under plane strain made of an isotropic elastic material characterized by its initial Young modulus $Y_0$ and Poisson ratio $\nu$. This domain, with a height-to-width ratio of 2, is discretized into an unstructured mesh using the finite element method \citep{amitrano1999}, i.e.~$N=960$ elements for a system size $L=16$. \HL{The height-to-width ratio is equivalent to that used in classical compression experiments on brittle materials like rocks or concrete in order to avoid the final fault formation from the corner of the specimen (which would be likely with a 1:1~height-to-width ratio). In the model, it avoids the final fault to occur through the diagonal most of the time, as the fault normal surface has typical values around $45-60^{\circ}$}. It is loaded under uniaxial compression along $y$, with the lower boundary remaining fixed along $y$ direction and the left and right boundaries allowed to deform freely~\citep{amitrano1999} to mimic the experimental setup of uniaxial compressive tests. In this athermal version, increasing the vertical shrinking along $y$ from the upper boundary simulates a strain-controlled loading. \HL{The local strain $\varepsilon_{n}$ is calculated using a usual finite element method procedure and the vertical strain at macroscale $\varepsilon$ is calculated from the average of the local vertical displacement in the upper boundary elements divided by the initial sample's height,}. When the stress locally exceeds, for one element $n$, a given threshold for damage, its elastic modulus $Y_{n}$ is multiplied by the damage factor $1-D$, with $D$ constant and small compared to 1 (set to 0.1 here). Each exceedance of the damage threshold induces a damage event in a given element $n$. After $\mathcal{D}_{n}$ accumulated damage events, the Young modulus of that element falls therefore to  $Y_{n}=(1-D)^{\mathcal{D}_{n}} Y_{n,0}$ \citep{girard2010}, with $Y_{n,0}$, the initial Young modulus of that element, while the Poisson's ratio is unaffected by this damage. After each damage event, the static equilibrium is recalculated, this way redistributing elastic stresses to neighbouring elements. The updated stress field is then compared to the local damage threshold for all elements. If some of these thresholds are exceeded, an avalanche of damage occurs. The avalanche stops when all the elements are below their threshold. The number of damage events in one loading step, i.e. while not increasing further the applied shrinking, defines the avalanche size \citep{amitrano1999}, which compares e.g.~with acoustic emissions recorded during compression tests on rocks (e.g.~Refs.~\citep{lockner1991quasi,lockner1993role,brantut2011damage,brantut2013}), concrete \citep{vu2019compressive}, or other brittle materials \citep{vasseur2015heterogeneity}. The local threshold for damage is expressed by a Mohr--Coulomb criterion, relevant, at a macroscopic level, for geomaterials under compressive stress states \citep{jaeger1979}, $\tau - \sigma\tan\phi = \tau-\sigma\mu=C$, where $\tau$ and $\sigma$ represent respectively the shear and the normal stress over an orientation that maximizes the Coulomb stress $\tau-\sigma\mu$, $\phi$ is the friction angle, $\tan\phi=\mu$ the corresponding internal friction coefficient, and $C$ the cohesion. This formulation of the Mohr-Coulomb criterion allows to compare two scalar stress values, the Coulomb stress \textit{vs} the cohesion and therefore, in the next section, to define a scalar stress gap. \HL{Nevertheless, the model is not scalar and the stress field variation could result in some elements failing by traction. For such rare cases we have implemented a traction criterion which compares a stress traction threshold to the minor stress, $C_t = \sigma_t - \sigma_3$}~\citep{amitrano1999, amitrano2006}
\HL{. This is compared with the one from Mohr--Coulomb criterion and the smallest stress gap chosen.}
To simulate a heterogeneous material, i.e.~to take into account microstructural heterogeneity \citep{hansen19904}, the cohesion $C$ is set randomly, and then kept unchanged to mimic quenched material heterogeneity, for each element from a Gaussian distribution, $C_n \sim \mathcal{N}(\langle C \rangle, \delta C^2)$, with arithmetic mean value $\langle C \rangle$ and standard deviation $\delta C$ (Fig. \ref{fig:pdm01}): 
\begin{eqnarray}\label{eqpdm002}
f(C_n) = \frac{1}{\delta C \sqrt{2\pi}} \exp \left[ -\frac{1}{2}\left(\frac{C_n-\langle C \rangle}{\delta C}\right)^2  \right],
\end{eqnarray}
while $\mu$ is set constant in space and time. However, as explained below, we performed two sets of simulations, one with $\mu=0.7$, a classical value for geomaterials \citep{jaeger1979}, another with $\mu=0$ for which the Coulomb criterion amounts to a Tresca plastic criterion independent of the pressure term, as observed for metals. This will allow to explore the role of the shape of the elastic interaction kernel, which depends on $\mu$ \citep{amitrano1999}.

Initially, the material is undamaged, $Y_{n,0} = Y_0$ $\forall  n$, and consequently the stress state is homogeneous. Once damage occurs, the stress field becomes heterogeneous, due to elastic stress redistribution. Athermal simulations are characterized by an initial linear elastic response before a macroscopic softening preceding a large stress drop mimicking a macroscopic rupture \citep{amitrano1999,girard2010}. This athermal model was shown to successfully represent Coulombic failure of disordered materials like rocks or concrete, e.g.~the progressive localization of damage upon approaching a peak stress at which an incipient fault nucleates \citep{lockner1991quasi}, or the impact of confining pressure and of the internal friction $\mu$ on strength and brittleness~\citep{amitrano2003}. Athermal simulations allow defining the athermal strength of a given sample from the peak stress reached before the macroscopic stress drop, $\sigma_{ath}$.

\begin{figure}[b!]
\includegraphics[width=\columnwidth]{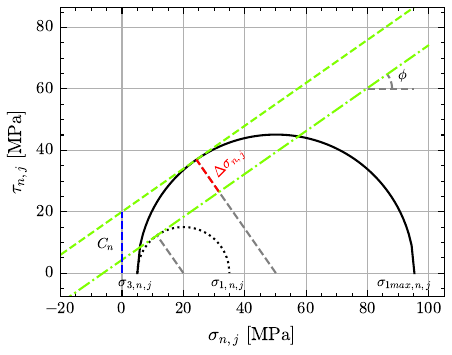}
\caption{\label{fig:pdm02} Representation of the Mohr--Coulomb criterion at microscale and the identification of its respective local stress gap $\Delta\sigma_{n,j}$.}
\end{figure}

\subsection{Thermal activation from a Kinetic Monte Carlo algorithm}

In the athermal, purely elasto-brittle version of the model, time is irrelevant and damage occurs only if the external stress and strain are increased. Therefore, it cannot describe creep deformation, i.e.~time-dependent deformation under constant load. In order to introduce thermal activation and a physical timescale, a BKL-KMC  algorithm~\citep{bortz1975, kratzer2009} was implemented \citep{weiss2023logarithmic}. At the local (element) scale, the gap $\Delta\sigma_n$ between the local stress state and the Coulombic envelope defines an energy barrier~$E_n$.
The goal of the KMC algorithm is to thermally activate damage events at the element scale according to the rates of the relevant individual processes. The jump rate of each element $n$ follows an Arrhenius expression~\citep{kratzer2009, bortz1975}:
\begin{eqnarray}\label{eqkmc001}
\omega_{n, j+1} = \Omega_0\exp\left[-\frac{E_{n,j}}{k_BT}\right],
\end{eqnarray}
where $E_{n,j} = V_a \Delta\sigma_{n,j}$ and $\Omega_0$ is a vibration frequency, which is, in solids, of the order of $10^{13}\; \si{\second}^{-1}$ \citep{zhurkov1965}. From Eq.~\ref{eqtheorApp001b}, the partition function at the $j$th transition is $Z_j=\Omega_0^{-1}\sum_n \omega_{n,j}$, and consequently the jump rate of the entire system is given by $\omega_j = \Omega_0 Z_j = \sum_n\omega_{n,j}$.

\begin{figure}[t!]
\centering
\includegraphics[width=\columnwidth]{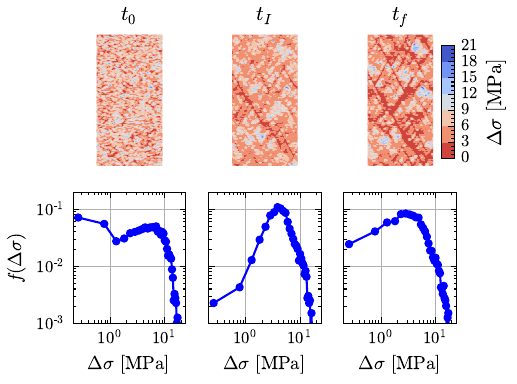}
\caption{\label{fig:pdm04} Example of the evolution of the stress gap field (top) and of the probability density $f(\Delta\sigma_j)$ of the stress gaps (bottom) for different thermal transitions $j$ at times: $t = 0$, end of primary creep $t = t_I$ and rupture time $t_f$ obtained from the numerical model.}
\end{figure}

In creep simulations, the model is initially launched athermally up to the targeted uniaxial applied stress,~$\sigma_{appl}$, defined as a fraction of the uniaxial strength~$\sigma_{ath}$ obtained by the athermal model for the same sample, with the same initial quenched heterogeneity. During this loading phase the model obeys athermal rules, so some elements can be damaged, possibly triggering avalanches. Creep starts from the end of this athermal loading, setting $t=0$, by thermally activating a first element $n$ from the KMC algorithm. This selection is done at random, weighted by the jump rates $\omega_{n,1}$ given by Eq.~\ref{eqkmc001}. Time advances by an increment $\Delta t_1$ following an exponential probability distribution as given in Eq.~\ref{eqtheor000c} and later iterates to obtain the following time increments $\Delta t_j \forall j \geq 1$. This time increment is obtained at any $j \geq 1$ thermal transition from Ref. \citep{bortz1975}:
\begin{equation}\label{eqkmc002}
\Delta t_j = - \log{u_j} \langle \Delta t \rangle_j = - \frac{\log{u_j}}{\Omega_0 Z_j},
\end{equation}
where $u_j \in [0,1]$ is a random number chosen from a uniform distribution.

The internal stress and strain fields are recalculated after the $jth$-thermal activation, potentially triggering an avalanche of damage events $\mathcal{D}_{n,j}$ while keeping the time unchanged, i.e.~the avalanche duration is considered to be negligible compared to the creep timescales. Once the avalanche stops, the KMC algorithm is re-launched to select the element to be thermally activated next, and to define the next time interval, and so forth. This is done according to updated individual jump rates (Eq.~\ref{eqkmc001}) resulting from modified stress gaps $\Delta\sigma_{n,j}$ after the elastic stress redistribution. 

\begin{figure}[tb]
\centering
\includegraphics[width=\columnwidth]{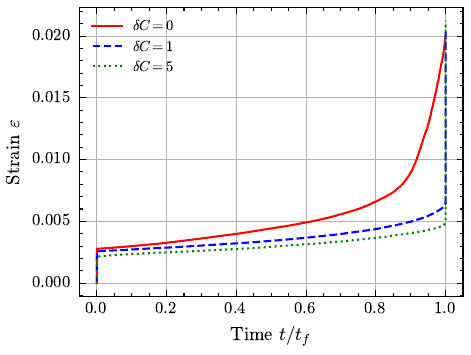}
\caption{\label{fig:pdm05} \HL{Example of the temporal evolution of the macroscopic axial strain for creep simulations with different heterogeneities.}}
\end{figure}

Different definitions of the stress gap $\Delta\sigma$ can be proposed, depending on the physical deformation mechanism. As an example, a von Mises stress gap has been proposed for amorphous plasticity \citep{castellanos2018}. In the present case, in agreement with our local Coulomb criterion and in order to simulate brittle creep \citep{scholz1968}, we used a Coulomb stress gap between a local cohesion $C_n$ and the local Coulomb stress, $\Delta\sigma_{n,j}\sim C_n-(\tau_{n,j}-\mu\sigma_{n,j})$, (Fig.~\ref{fig:pdm02}). Its expression at the local scale reads:
\begin{eqnarray}\label{eqkmc003}
\Delta\sigma_{n,j}=C_n\cos\phi + \HL{\sigma_{3,n,j}}\frac{1+\sin\phi}{2} - \sigma_{1,n,j}\frac{1-\sin\phi}{2},
\end{eqnarray}
where the subindex \HL{$_{n,j}$ indicates the element and the transition respectively}, $C_n$ is the local cohesion, $\sigma_{1, n,j}$ and $\sigma_{3, n, j}$ are respectively the local major and minor principal stresses, and $\phi$ is the internal friction angle. \HL{The intermediate principal stress under the plane strain hypothesis corresponds to $\sigma_{2, n, j}=\nu(\sigma_{1, n, j} + \sigma_{3, n, j})$. The equation} \ref{eqkmc003} \HL{can also be written in Cartesian coordinates doing the following variable change:}
\begin{multline}\label{eqkmc003b}
\HL{
\sigma_{1,n,j},\sigma_{3,n,j}=\frac{\sigma_{x,n,j}+\sigma_{y,n,j}}{2}} \\ \HL{\pm \sqrt{\left(\frac{\sigma_{x,n,j} -\sigma_{y,n,j}}{2}\right)^2 + \tau_{xy,n,j}^2},
}
\end{multline}
\HL{where $\sigma_{x,n,j}$ and $\sigma_{y,n,j}$ are the local horizontal and vertical stresses respectively, and $\tau_{xy,n,j}=\tau_{yx,n,j}$ are the local shear stresses in the horizontal and vertical plane.}

The probability distribution of the stress gaps, 
$f(\Delta\sigma)$, which evolves during the test (Fig.~\ref{fig:pdm04}), corresponds to a convolution of probability distributions \citep{sornette2006}. It can be written as:
\begin{multline}\label{eqkmc004}
f(\Delta\sigma_j) = \left( f_C \ast f_{\sigma_1, \sigma_3} \right) (\Delta\sigma_j) \\
\sim \int_{-\infty}^{\infty} f_C(C\cos\phi) f_{\sigma_1, \sigma_3}(\Delta\sigma_j - C\cos\phi) dC
\end{multline}
with $f_C = f(C \cos\phi)$ the probability distribution of the cohesion term, set as a Gaussian distribution in our setup, and $f_{\sigma_1, \sigma_3}$ the probability distribution of the stress state, $f_{\sigma_1, \sigma_3}(\Delta\sigma - C\cos\phi) = \left( f_{\sigma_1} \ast f_{\sigma_3} \right) (\Delta\sigma - C\cos\phi)$.

If during the initial athermal stage no damage occurs, the only term affecting the probability distribution of the stress gaps at the onset of creep is $f_C$, as the initial stress field is homogeneous  and the principal stresses $\sigma_1, \sigma_3$ are spatially invariant. Consequently, the stress gap probability distribution at $t=0$ is also Gaussian, comparable to the theoretical expressions considered in Section \ref{chap_analyt}. \HL{In the model, final failure is assumed to occur, and so the simulation terminates, once the axial macroscopic strain $\varepsilon_{j} \geq \varepsilon_{max}$, with $\varepsilon_{max}\approx 2\times10^{-2}$, thus defining the failure time, $t_f$. In all cases, this happens during a tertiary creep regime for which the strain rate eventually diverges, $\dot{\varepsilon}_{j} \to \infty$ (see Fig. \ref{fig:pdm05}), just like in experiments.}

\section{Summary of the used parameters and their corresponding notations}\label{app:d}

Table \ref{t1} shows the parameters used in this paper. It includes the symbols, descriptions, values in cases where it is applicable, and the units.

\begin{table*}
\begin{center}
\begin{tabular}{||c c c||} 
 \hline
 Symbol & Description & Value, unit  \\ [0.5ex] 
 \hline\hline
 $C_n$                        & local cohesion       & \si{\mega\pascal} \\
 $\langle C \rangle$          & arithmetic mean value of the cohesion & 20~\si{\mega\pascal} \\
 $D$                          & damage parameter                    & \num{0.1}    \\
 $\mathcal{D}_n$              & local cumulated damaged events (due to avalanches) during the athermal load ($\mathcal{D}_{n,j}$ during creep) & - \\
 $E$                          & activation energy    & \si{\joule}  \\
 $E_{n,j}$                    & local activation energy at event (transition) $j + 1$ & \si{\joule}\\
 $E_{\Delta\sigma_h,0}$       & arithmetic average of the activation energy for the first event (transition)                & \si{\joule}       \\
 $\langle E \rangle_{j}$      & average energy weighted by the activation energy probability $p(E_{n,j})$, at the event (transition) $j + 1$ & \si{\joule} \\
 $\langle E_{\Delta\sigma} \rangle_{j}$  & arithmetic average of the activation energy at the event (transition) $j + 1$  & \si{\joule}   \\
 $f(E_{n_b,j})$               & relative density of states  &   -       \\
 $F_j$                        & free energy  & \si{\joule}       \\
 $F_{h,1}$                    & free energy (from an "equivalent" homogeneous system) at the first event (transition) & \si{\joule}       \\
 $f(\Delta\sigma_{n_b,j})$    & probability distribution of the stress gap  &   -   \\
 $g(E_{n_b,j})$               & density of states (number of $n_b$ sub-volumes with energy within a small range $(E_{n_b,j},E_{n_b,j} + dE_{n_b,j}$)) &   -    \\
 $j$                          & index indicating a thermally activated event (transition) &-\\
 $k_B$                        & Boltzmann's constant  &\num{1.38e-23} \si{\joule\kelvin}$^{-1}$ \\
 $L$                          & system size      & 16  \\
 $n$                          & index indicating a sub-volume (micro-element)  & -  \\
 $N$                          & total number of sub-volumes forming a solid volume   & - \\
 $\mathcal{N}_f$              & number of successive thermally activated events (transitions) leading to failure  & -      \\
 $p(E_{n,j})$                 & probability of a solid composed of $N$ sub-volumes to be in a configuration with energy $E_={n,j}$ & -  \\
 $S_{j}$                      & Shannon's entropy & \si{\joule\kelvin}$^{-1}$  \\
 $T$                          & absolute temperature     & \si{\kelvin}  \\
 $T_{eff}$                    & effective temperature      & \si{\kelvin}  \\
 $t_f$                        & rupture time      & \si{\second} \\
 $\langle t_f \rangle$        & average rupture time for different realizations of the initial quenched heterogeneity  & \si{\second}\\
 $\langle t_f \rangle^{gauss}$& predicted average rupture time from a Gaussian distribution of the initial quenched heterogeneity   & \si{\second}\\
 $V_{a}$                      & activation volume for the damage / fracturing microscopic mechanism         &\num{11e-27}~\si{\meter}$^{3}$   \\
 $V_{a,est}$                  & estimated activation volume for the damage / fracturing microscopic mechanism   & -   \\
 $Y_0$                        & Young' modulus of reference & 21 \si{\giga\pascal}\\
 $Y_n$                        & local Young' modulus after a damage event occurs during the athermal load ($Y_{n,j}$ during creep) & \si{\giga\pascal}\\
 $Z_{j}$                      & partition function at the event (transition) $j + 1$ & -   \\
 $\delta C$                   & standard deviation of the cohesion       & 0,1,5 \si{\mega\pascal} \\
 $\delta E_{\Delta\sigma,j}^2$& variance of the activation energy at the event (transition) $j + 1$  & \si{\joule}$^2$   \\
 $\delta_{\Delta\sigma,j}^2$  & variance of the stress gap at the event (transition) $j + 1$                      &   \si{\mega\pascal}$^2$ \\
 $\Delta F_j$                 & differential free energy  & \si{\joule}     \\
 $\Delta t_{h,1}$             & "equivalent" initial time step (from an "equivalent" homogeneous system)   & \si{\second}\\
 $\Delta t_j$                 & waiting time between events (transitions) & \si{\second}\\
 $\overline{\langle\Delta t\rangle_{j}}$ & arithmetic mean value of all the time intervals between successive events (transitions) & \si{\second}\\
 $\langle\Delta t\rangle_j$   & average waiting time between events (transitions) & \si{\second}\\
 $\Delta\sigma_{macro}$       & stress gap at macroscale between a (supposedly known) athermal strength and the applied stress  & \si{\pascal}       \\
 $\Delta\sigma_{n,j}$         & local stress gap at event (transition) $j + 1$  & \si{\pascal} \\
 $\langle \Delta\sigma \rangle_{j}$      & arithmetic average of the stress gap at the event (transition) $j + 1$     & \si{\mega\pascal}   \\
 $\sigma_{appl}$              & applied creep stress at macroscale  & \si{\mega\pascal}      \\ 
 $\sigma_{ath}$               & macro athermal threshold  & \si{\mega\pascal}      \\
 $\sigma_{ath,n}$             & local athermal threshold          & \si{\pascal}       \\
 $\sigma_{n,j}$               & local stress state at event (transition) $j + 1$  & \si{\pascal}\\
 $\sigma_{appl}/\sigma_{ath}$ & stress ratio (percentage of the athermal strength)     & -    \\
 $\mu=\tan\phi$               & internal friction coefficient (function of the internal friction angle $\phi$)        & 0.0, 0.7 \\
 $\nu$                        & Poisson's ratio                      & 0.25   \\
 $\Omega_0$                   & thermal vibration frequency     &\num{1e13}~\si{\second}$^{-1}$\\
 $\omega_j$                   & jump rate at event (transition) $j$  &\si{\second}$^{-1}$\\
 [1ex] 
 \hline
\end{tabular}
\caption{\label{t1} Summary of the notation of the different parameters used.}
\end{center}
\end{table*} 

\end{document}